\documentclass[10pt]{article}
\usepackage[utf8]{inputenc}

\usepackage{parskip}
\usepackage[margin=1in]{geometry}
\usepackage{amsmath}
\usepackage{amssymb}
\usepackage{graphicx}
\usepackage{xcolor}
\usepackage{longtable}
\usepackage{enumerate}
\usepackage[lined,boxed]{algorithm2e}
\usepackage{cmbright}
\usepackage{natbib}
\usepackage{float}
\usepackage{algorithm2e}
\usepackage{arydshln}

\newcommand{\data}[1]{\textbf{Data Acknowledgement:} #1}
\newcommand{\funding}[1]{\textbf{Funding Acknowledgement:} #1}
\newcommand{\keywords}[1]{\textbf{Keywords:} #1}

\usepackage{setspace}

\newcommand{\bs}[1]{\boldsymbol{#1}}
\newcommand\ci{\perp\!\!\!\perp}

\makeatletter
\renewcommand{\maketitle}{\bgroup\setlength{\parindent}{0pt}
\begin{flushleft}
  \LARGE
  \textbf{\@title}
  
  \large
  \@author
  
  \large
  \@date

\end{flushleft}\egroup
}
\makeatother

\title{Nonparametric Estimation of Population Average Dose-Response Curves using Entropy Balancing Weights for Continuous Exposures}
\author{Brian G. Vegetabile, RAND Corporation \hfill correspondence: bvegetab@rand.org \\ 
        Beth Ann Griffin, RAND Corporation \\ 
        Donna L. Coffman, Temple University \\ 
        Matthew Cefalu, RAND Corporation \\ 
        Daniel F. McCaffrey, Educational Testing Service\\ 
        \vspace{1em}}
\date{March 2020}

\begin{document}

\maketitle

\begin{abstract}
Weighted estimators are commonly used for estimating exposure effects in observational settings to establish causal relations. These estimators have a long history of development when the exposure of interest is binary and where the weights are typically functions of an estimated propensity score.  Recent developments in optimization-based estimators for constructing weights in binary exposure settings, such as those based on entropy balancing, have shown more promise in estimating treatment effects than those methods that focus on the direct estimation of the propensity score using likelihood-based methods.  This paper explores recent developments of entropy balancing methods to continuous exposure settings and the estimation of population dose-response curves using nonparametric estimation combined with entropy balancing weights, focusing on factors that would be important to applied researchers in medical or health services research. The methods developed here are applied to data from a study assessing the effect of non-randomized components of an evidence-based substance use treatment program on emotional and substance use clinical outcomes.  \newline

\end{abstract}



\section{Introduction}

Assessing causal relationships between continuous treatments and clinical outcomes is often of interest in substance use research. For example, we may seek to understand how substance use outcomes are affected by the dose of a substance use treatment program. Evaluations of such programs can be analytically challenging because researchers are often unable to control the factors (observed and unobserved) that would allow them to fully understand the relationship between the treatment program and the clinical outcomes. The gold standard for making causal claims is a randomized experiment in which control over important variables would be attained through randomization.  Use of observational studies in health care domains, such as in our motivating example assessing the effect of non-randomized components of an evidence-based substance use treatment program on clinical outcomes, offer more difficult settings with which to assess causal relationships. Strong assumptions must be made to defend claims that would be considered even close to causal in these settings. Statistical methods from the field of causal inference provide much needed tools that enable researchers to come closer to addressing the questions they would hope to address using a clinical trial, assuming such strong assumptions can be asserted. 


The field of causal inference has an extended history for addressing the challenges inherent in observational studies, often focused on methods involving the propensity score \citep{rosenbaum_rubin_1983} and settings where the exposure of interest is binary (see, \cite{rosenbaum2010design,imbens_rubin_2015} for examples and extended history).  More recent extensions of these methods to settings where the exposure of interest is not binary have arisen, and are often focused on the estimation of \textit{population dose-response curves} \citep{robins_hernan_brumback_2000,imbens2000,hirano2004propensity,imai_vandyk_2004} that describe the relationship between a continuous treatment and an outcome.  Many of these extensions in observational data settings estimate the dose-response curve conditional on the ``generalized propensity score'' (GPS) through stratification or inverse probability weighting (IPW). Unfortunately, these methods appear to be under-utilized in the medical and health care literature, potentially because there is not enough guidance for the research community to readily adopt these methods and/or because their use requires asserting strong assumptions 
that may be unlikely to hold in practice.  An assumption of the correct specification of the propensity score is difficult enough to assert in the binary setting; the assumption of the correct \textit{distribution} of the GPS needed in estimating the dose-response curve in the continuous treatment setting can be even more difficult.  

There has been some work to improve the utility of methods for continuous treatments or exposures by focusing on flexible modeling of the GPS and outcome models \citep{zhu2015boosting,kennedy2017}.  Unfortunately, flexible models can sometimes be insufficient to fully reduce bias in observational studies when models are misspecified (e.g., the GPS requires modeling the entire distribution of the exposure given observed covariates and misspecification can occur if flexible modeling is used to only estimate the mean relationship between exposure and covariates).  It may be that there is more promise, as has been shown in the binary settings, when using optimization based weighting methods that seek to directly optimize ``covariate balance''  \citep{hainmueller2012entropy,imai_ratkovic_2014,zubizarreta2015stable,vegetabile2020}. These methods remove the requirement of modeling the propensity score or GPS directly.  One such method, entropy balancing, has recently been demonstrated to have properties similar to doubly robust estimators \citep{zhao2017entropy}, where if either the outcome model or assignment model is specified correctly, then consistent estimation is possible.  These advances in optimal balancing have begun to be applied to continuous exposure settings \citep{yiuli2018,fong2018covariate,kallus2019kernel} and recent work has extended the entropy balancing methodology to this setting \citep{tubbicke2020entropy}.  These methods hold great promise for the estimation of more accurate dose-response curves in health care and medical research, and a richer understanding of dose-response curves has the potential to save lives and improve patient outcomes.

In this paper, we examine the use of entropy balancing for continuous treatments and illustrate how it can be used with nonparametric regression methods for estimating population dose-response curves. Our work is distinct from concurrent research on this topic \citep{tubbicke2020entropy} in the following ways: (1) we consider nonparametric methods for the estimation of the dose-response curve that better reflect real world relationships; (2) we provide a careful discussion of the influence that higher order moments of the exposure and covariates have on the entropy balancing algorithm and subsequent estimation of the dose-response curve; (3) we provide evidence on the performance of bootstrap-based confidence intervals for point-wise inference of the dose-response curve; and (4) we include a user-friendly R package for deriving entropy balancing weights.  Throughout the paper, we emphasize considerations that must be made in practice in order to provide researchers with needed guidance on implementing these methods. Our motivating case study examines the causal effects of the number of treatment components received for an evidence based substance use treatment program for adolescents on both substance use and emotional well-being outcomes. The use of nonparametric estimation and entropy balancing together removes the need for the researcher to choose a correct model specification of the GPS and a correct form of the outcome model, removing two key sources of potential error in analyses. 

The organization of the paper is as follows: Section \ref{sec:methods} provides the development of the methodology, including an overview of causal inference in the potential outcomes framework for continuous exposures, the development of entropy balancing for continuous exposures, careful consideration of key issues for applied analysis (i.e., the assessment of covariate balance and choosing the number of moments to balance), and details on inference, the local linear regression strategy, and bootstrapped standard error estimators.  Section \ref{sec:simulation} provides a simulation that attempts to stress the method in reasonable ways (e.g., non-linearity in the exposure/covariate relationships and non-linear outcome models). Section \ref{sec:application} applies the methods to data collected from over 2200 adolescents substance users receiving the Adolescent Community Reinforcement Approach (A-CRA) \citep{godley2016adolescent} treatment program with the goal of assessing the causal effect of the number of behavioral treatment components received as part of their A-CRA program, on longitudinal substance use and emotional well-being outcomes. Section \ref{sec:discussion} provides a concluding discussion and remarks on the use of these methods for health services researchers.  

\section{Methods} \label{sec:methods}

In this section, we provide an overview of the methods for nonparametric estimation of population average dose-response curves using entropy balancing.  We begin with an overview of causal inference in the continuous exposure setting that introduces our notation throughout.  We then provide an extension of entropy balancing for continuous exposure settings and demonstrate how these weights can be used with local regression estimation techniques that can incorporate the entropy balancing weights and nonparametrically estimate dose-response curves.  As stated in the introduction, we have developed these methods into an \texttt{R} package (project development can be found at https://github.com/bvegetabile/entbal).  

\subsection{Notation and Causal Inference Framework} \label{sec:causalframework}

We utilize a generalization of the potential outcomes framework of \cite{neyman1923} and \cite{rubin1974}, specifically focusing on notation from the extension to general exposure settings outlined in \cite{hirano2004propensity}.  We define an index set $i \in \{1, \dots, N\}$ of units for that we are interested in assessing some ``dose-response'' relationship, i.e., a functional relationship between an exposure and an outcome within some population.  We define the exposure variable for each unit as $A_i$ taking values from the set $\mathcal{A}$, e.g., $A \in \Re^{+}$, and define a random vector $\bs{X}_i \in \mathcal{X}$ as the set of variables that occur prior to the application of the exposure $A_i$ and are responsible for setting its value; typically $\bs{X}_i=(X_{i1}, X_{i2}, \dots, X_{ij}, \dots, X_{id})$ will be a $d$-dimensional vector and embedded in $\Re^{d}$. We also define the observed outcome of interest as a variable $Y_i$ and we posit that \textit{for each unit} there exists a potential outcome $Y_i(a)$ for all $a \in \mathcal{A}$, where $Y_i = Y_i(a)$ if $A_i = a$ and $Y_i(a')$ is unobserved for all other $a' \in \mathcal{A}$.  

Under this framework, an ideal measurement process would observe an infinite number of copies of each unit, apply the exposure at exactly the same instance in space and time across these copies and measure each outcome $Y_i(a)$; this is clearly not possible in most practical cases and in the case where there are only two exposures, this has been referred to as the \textit{Fundamental Problem of Causal Inference} \citep{holland1986}.  Therefore, we leverage information across units and estimate the functional $f(a) = E[Y(a)] \equiv \int_{\mathcal{X}} Y(a) p(\bs{x}) \partial \bs{x}$, where the expectation is with respect to a population of individuals defined by the random vector $\bs{X}$. The functional $f(\cdot)$ defines the \textit{population} dose-response curve, i.e., the expected response to the application of $A=a$ in a population defined by the vector $\bs{X}$. Using this functional, we could choose to define causal effects as being contrasts of $f(a)$, e.g., $\tau(a,a') = f(a') - f(a) = E[Y(a') - Y(a)]$.

There are numerous estimation strategies for the estimation of $f(\cdot)$ within the Neyman/Rubin framework, e.g., regression estimation \citep{hirano2004propensity} or through the use of subclassification on the GPS \citep{imai_vandyk_2004}.  In this paper, we utilize weighted estimation similar to the marginal structural approach \citep{robins_hernan_brumback_2000}.  The core of the procedure is the result that if we define 
\begin{align}
    w_i \equiv w_i(a, \bs{x}) = \frac{\mbox{Pr}(A_i=a)}{\mbox{Pr}(A_i=a | \bs{X}_i= \bs{x})} =  \frac{p(a)}{p(a|\bs{x})}, \hspace{1em} \mbox{ where } p(\cdot) \mbox{ denotes a density function,} \label{eqn:weights}
\end{align}
then we can obtain consistent estimation for the average response, i.e., it follows that 
\begin{align}
    E[w_i Y | A = a ] = E[Y(a)]. \label{eqn:consistent}
\end{align}
Note that the denominator of Equation \ref{eqn:weights}, i.e., $e(a,\bs{x}) \equiv p(a|\bs{x})$, is referred to as the GPS in the literature and the weights are referred to as stabilized inverse probability of treatment weights (IPTW). 

The results in Equation \ref{eqn:consistent} require a few strong assumptions (see \cite{imai_vandyk_2004} and \cite{hirano2004propensity} for overviews) for consistent estimation and a causal interpretation, specifically: 1) an extension of the \textit{stable unit treatment value assumption} (SUTVA) for continuous exposure settings, and 2) ignorability assumptions. Assumption 1 states that the distributions of the potential outcomes for a unit are assumed independent of the potential exposure status of another unit, given observed covariates.  This assumption implies that there is no interference between units and that we can use a consistency argument to assert that the observed outcome under treatment $A=a$ is equal to the corresponding potential outcome, i.e., $Y_i = Y_i(A_i)$.  The ignorability assumptions contain two requirements: 1) unconfoundedness, and 2) positivity.  Unconfoundedness requires that the potential outcomes are conditionally independent of treatment given the covariates, e.g., a weak unconfoundedness assumption defined in \cite{hirano2004propensity} states that $Y(a) \ci A \ | \ \bs{X}$ for all $a\in\mathcal{A}$.  Ignorability assumptions also require \textit{positivity}, i.e., $0 < p(A \in \mathcal{T} | \bs{X})$ for all $\bs{X} \in \mathcal{X}$ and measurable sets $\mathcal{T}\subset \mathcal{A}$ with positive measure \citep{imai_vandyk_2004}.  The implication of the ignorability assumptions are that \textit{all} covariates responsible for assigning exposure levels have been identified and any missing or unobserved $\bs{X}^*$ variable is a potential source of estimation bias.  They also imply that for any given unit with $\bs{X}=\bs{x}$ there needs to be sufficiently high probability of receiving any treatment level under consideration.  These are both \textit{strong} assumptions, e.g., unconfoundedness is untestable \citep{imbens_rubin_2015}, that require thought and defense in any analysis. Additionally, positivity may not hold if it is observed that there is a strong relationship between any covariate and the exposure.  For this paper, we assume that these assumptions hold.   

Using these assumptions, we will use weighted estimation methods to estimate the population dose-response curve and develop a method for constructing optimal weights.   

\subsection{Entropy Balancing for Continuous Exposures} \label{sec:eb}

Entropy balancing \citep{hainmueller2012entropy} is a method of creating optimal weights for causal estimation.  Within the causal literature, the method was originally developed for the estimation of causal effects in binary exposure settings, though related methods have a longer history in survey calibration \citep{haberman1984adjustment,deville1992calibration,deville1993generalized}.

Here we demonstrate an extension of entropy balancing to handle continuous exposures. This has been proposed in \cite{tubbicke2020entropy}, but we adopt a slightly different notation to provide guidance on the influence that higher moments of the exposure and covariates have in the entropy balancing algorithm and the subsequent estimation of dose-response curves. If we define $\mu_j^p = E[X_j^p]$, the mean of an element $X_j$ from vector $\bs{X}$, and $\mu_A^q = E[A^q]$ for integers $p$ and $q$ and assume that we have the correct weights as defined in  (\ref{eqn:weights}), then the method relies on the observation that in the joint distribution of $A$ and $\bs{X}$, the weighted covariance between $X_j^p$ and $A^q$ for any pair $(p,q)$ should be zero, while all other weighted marginal relationships should remain the same.  This is demonstrated for the covariance between $\bs{X}$ and $A$ here, 
\begin{align}
    E[w(X_j - \mu_j)(A - \mu_A)] &= \int_{\mathcal{A}} \int_{\mathcal{X}} w (x_j - \mu_j)(a - \mu_A) p(a, \bs{x}) \partial a \partial \bs{x} \label{eqn:integration} \\
    &= \int_{\mathcal{A}} \int_{\mathcal{X}} \frac{p(a)}{p(a|\bs{x})} (x_j - \mu_j)(a - \mu_A) p(a, \bs{x}) \partial a \partial \bs{x} \nonumber\\
    &= \int_{\mathcal{A}} \int_{\mathcal{X}} \frac{p(a) p(\bs{x})}{p(a|\bs{x})p(\bs{x})} (x_j - \mu_j)(a - \mu_A) p(a, \bs{x}) \partial a \partial \bs{x} \nonumber\\
    &= \int_{\mathcal{A}} \int_{\mathcal{X}} (x_j - \mu_j)(a - \mu_A) p(a)p(\bs{x}) \partial a \partial \bs{x} \nonumber\\
    &= 0 \nonumber
\end{align}
This observation will allow us to select nonparametric weights, i.e., those that do not model $p(a)$ or $p(a|\bs{x})$ directly, that enforce the constraints that we would expect to observe if the weights had the correct ratio-of-densities interpretation.  A similar constraint was utilized in \cite{fong2018covariate} for their nonparametric covariate balancing propensity score (npCBPS) for generalized treatment regimes.

To select weights, we will minimize the Kullback-Leibler divergence between two distributions of weights: one that utilizes the weights as defined above and another based on empirical weights, i.e., $q_i \equiv N^{-1}$, subject to constraints on weighted covariance between $A$ and $\bs{X}$ and enforcing constraints on the marginal distributions.  This minimization has the effect of finding a set of weights $w$ that are ``closest'' in the sense of KL-divergence to a set of weights that maximize the sample information.  We will also impose additional constraints that ensure the weights are positive and sum to one. This is formalized below:
\begin{center}
\small
\begin{tabular}{rlrr}
    $\displaystyle\min_{w} H(w,q) = $ & $\displaystyle\sum_{i =1}^N w_i \log \left( \frac{w_i}{q_i} \right)$ & \hspace{1em} subject to   \\
   1) &$\displaystyle \sum_{i =1}^N w_i (X_{ij}^p - \mu_{j}^p)(A_i - \mu_A) = 0$ & \hspace{1em} for each $j$ and $p=1,2,\dots$ & (covariance constraints) \\ 
   2) &$\displaystyle \sum_{i =1}^N w_i (A_i^q - \mu_A^q) = 0$ & \hspace{1em} for specified $q=1,2,\dots$ & (marginal constraints: $A$) \\ 
   3) &$\displaystyle \sum_{i =1}^N w_i (X_{ij}^p - \mu_{j}^p) = 0$ & \hspace{1em} for each $j$ and $p=1,2,\dots$ & (marginal constraints: $X$) \\ 
   4) &$\displaystyle \sum_{i =1}^N w_i = 1, \hspace{1em} w_i \ge 0$ && (weight constraints)
\end{tabular}
\end{center}

This can be optimized using the method of Lagrange multipliers, where the objective becomes
\begin{align}
    \min_{w, \lambda, \gamma, \phi} & \left[\sum_{i =1}^N w_i \log \left( \frac{w_i}{q_i} \right) + \sum_{j,p} \lambda_{j,p} \sum_{i =1}^N w_i (X_{ij}^p - \mu_{j}^p)(A_i - \mu_a) \right. \nonumber \\ 
    & \hspace{1em} \left. + \sum_{q} \gamma_{q} \sum_{i =1}^N w_i (A_i - \mu_a) + \sum_{j,p} \phi_{j,p} \sum_{i =1}^N w_i (X_{ij}^p - \mu_{j}^p) + (\lambda_0 - 1) \left(\sum_{i=1}^N w_i - 1\right)\right] \label{eqn:objective}
\end{align}
The solution to this objective is of a similar form to that found in \cite{hainmueller2012entropy}, but here the weights are of the form,
\begin{align*}
    w_i = \frac{q_i \exp\left(- \left[\sum_{j,p} \lambda_{j,p} (X_{ij}^p - \mu_{j}^p)(A_i - \mu_a) + \sum_{q} \gamma_{q} (A_i^q - \mu_a^q) + \sum_{j,p} \phi_{j,p} (X_{ij}^p - \mu_{j}^p) \right] \right)}{\sum_{i=1}^N q_i \exp\left(- \left[\sum_{j,p} \lambda_{j,p} (X_{ij}^p - \mu_{j}^p)(A_i - \mu_a) + \sum_{q} \gamma_{q} (A_i^q - \mu_a^q) + \sum_{j,p} \phi_{j,p} (X_{ij}^p - \mu_{j}^p) \right] \right)}.
\end{align*}
Plugging these weights into Equation (\ref{eqn:objective}) we find the dual form that is no longer a function of the weights,
\begin{align*}
    \min_{\lambda,\gamma,\phi} \log\left( \sum_{i=1}^N q_i \exp\left(- \left[\sum_{j,p} \lambda_{j,p} (X_{ij}^p - \mu_{j}^p)(A_i - \mu_a) + \sum_{q} \gamma_{q} (A_i^q - \mu_a^q) + \sum_{j,p} \phi_{j,p} (X_{ij}^p - \mu_{j}^p) \right] \right) \right)
\end{align*}
This new dual objective can now be optimized using an efficient convex optimization routine and has a reduced parameter space.  Our implementations have found the limited memory Broyden–Fletcher–Goldfarb–Shanno algorithm with bounding constraints (L-BFGS-B) implemented in \texttt{R} to be computationally efficient.  The above optimization is in some cases numerically unstable, i.e., computational overflows can occur when exponentiating the sum of the multiplications of $\lambda, \gamma, \phi$ and powers of each observation's constraint variables (e.g., $X^p_{ij} - \mu_j^p$), and therefore we further constrain the optimization to be bounded such that parameters lie in the range $\lambda, \gamma, \phi \in [l, u]$.  In practice, these are set to large values, e.g., $[l, u]  = [-100,100]$ and can be further constrained if there are numerical instabilities. Note that in practice when these numerical issues occur, there is often an issue in satisfying the optimization constraints. The bounding constraints are used in our implementation to ensure that the algorithm at least returns a set of weights to the user that can be evaluated.  This behavior though implies that ``perfect'' balance with respect to the constraints is not possible in this data, which may be observed by remaining imbalances when performing covariate balance assessments, or if the $\lambda, \gamma, \phi$ are equal to the constraint values.  When these parameters must be constrained to lie in a small range in practice, it should be assessed how sensitive the results are to this bounding choice.

For more details on the original formulation of this problem in binary settings, see \cite{hainmueller2012entropy} and Appendix \ref{app:binary} for an overview. 

\subsubsection{Choosing the number of moments to balance}

The above algorithm requires specifying the number of moments, i.e., the parameters  $(p,q)$, that must be balanced. As suggested by Equation (\ref{eqn:integration}), the definition of the weights implies that the marginal distributions of $A$ and $X$ should remain unchanged under weighting and that only the covariances should be zero. Clearly though, as more constraints are placed on the optimization procedure it is more challenging to find a solution set and typically the effect of additional moment constraints is a reduction in the ``effective'' number of usable observations for analysis (discussed further in Section 2.2.2).  To keep the marginals correct for $A$ and $\bs{X}$, we have observed that in our simulations balancing two or three moments when variables are continuous and one moment when they are binary (ordinal variables are treated as continuous and multi-valued variables are converted to binary variables through dummy coding) has provided adequate performance. The higher moment, i.e., $p,q=3$, may be helpful in cases where the treatment or covariates are skewed, or if there are non-linearities in the relationship between $A$ and elements of the vector $\bs{X}$.  Similarly, we do not remove covariances between $X_j^p$ and $A$ for $p > 3$, as typically such strong relationships would imply that the positivity assumption is violated, or the effects of higher moments are weak enough that they can be captured with lower moments.  

\subsubsection{Assessing weighted balance and the effect of weighting on sample information}

An important step in any causal analysis is assessing ``covariate balance'' and the reduction in sample information due to the weighting scheme employed (captured by the effective sample size in analyses that utilize the weights). Covariate balance in the binary setting is typically assessed by comparing both unweighted and weighted standardized mean differences or other test statistics (e.g., Kolmogorov–Smirnov statistics) to understand how well the weights have improved the overall comparability of the two groups.  Visual inspection is also often performed comparing the weighted empirical cumulative distribution function (CDF) to the CDF of the target population for inference for the covariates.  The idea behind these balance assessments is that when the weights are applied to each group (e.g., exposure versus control) the weighted distribution of the pre-exposure confounders will be similar to the target population over which inference is being estimated (e.g., for the average treatment effect in the population (ATE) the target is the overall population characteristics; for the average treatment effect among those that are treated, the target is the characteristics of the treatment group). 

In the continuous exposure setting, we will perform similar diagnostics from the binary setting, but include more comparisons to ensure we have adequately assessed covariate balance.  Note that these balance assessments are not specific to entropy balancing weights and would apply for any type of weights that may be used in a continuous exposure setting, but throughout we focus on how they apply in our setting.  A comprehensive assessment of covariate balance should focus on the following summaries between the weighted and unweighted distributions:
\begin{itemize}
    \item \textit{Conditional mean balance}: linearly regressing $A$ on each $X_j$ and assessing the model coefficient to provide the magnitude of the relationship between the two variables
    \item \textit{Correlation balance}: assessing the Pearson correlation between $A$ and each $X_j$
    \item \textit{Marginal balance}: assessing marginal summary statistics of the variables in the data set, i.e., each $X_j$ and $A$, specifically calculating the following summaries:
    \begin{enumerate}
        \item Marginal mean
        \item Marginal variance
        \item Kolmogorov–Smirnov statistics
    \end{enumerate}
    \item \textit{Effective Sample Size}: summarizing the number of observations in the each sample
\end{itemize}

The conditional mean balance is a similar assessment as in the binary setting, i.e., we are leveraging the fact that if our weights have the correct ratio-of-density interpretations then $E[wX_j|A=a] = \int_{\mathcal{X}}w x_j p(\bs{x} | a) \partial \bs{x} = \int_{\mathcal{X}} x_j \frac{p(a)p(\bs{x})}{p(a, \bs{x})}  \frac{p(a,\bs{x}) }{p(a)} \partial \bs{x} = \int_{\mathcal{X}} x_j p(\bs{x}) \partial \bs{x} = E[X_j]$.  This observation combined with our constraint that the covariance must be zero imply that the model coefficient in a weighted regression using our entropy balancing weights should also be zero if the constraints can be satisfied.  The conditional mean balance is different than the binary setting where we do not perform a regression and simply compare the weighted means in each group, i.e., comparing $E[wX | A= a]$ and $E[wX|A=a']$ to see if they are the same.  

The next assessments, correlation balance and marginal balance, investigate if the weights have achieved zero correlation between the exposure and each covariate while maintaining the original marginal distributions over which inference is required. One concern would be that if the weighted exposure distribution is different than the unweighted distribution, then it implies that the weights have changed the distribution of the exposure over which inference is possible. To assess correlation balance we leverage the fact that the primary constraint in the optimization was that the weighted covariance between each $X_j^p$ and $A$ is zero. To fairly compare the performance across covariates on a standardized scale we instead look at the Pearson correlation coefficient between $X_j^p$ and $A$ in both the weighted and unweighted distributions. If the optimization is successful, then these values should be very small (up to the tolerance of the optimization routine), though it has been suggested elsewhere that when using the GPS, values less than 0.1 should be acceptable in practice \citep{zhu2015boosting}. The optimization also specifies that certain moments of the marginal distributions of the exposure and covariates are unchanged in the weighted distribution.  To validate this, we compare the unweighted and weighted mean and variance of the exposure variable and covariates to assess that they are unchanged by the weighting. Similarly, the KS-statistic provides an additional assessment of the distortion of the marginal distribution and we compare the KS-statistics between the weighted and unweighted distributions.  As the algorithm does not balance all moments (typically we set the max $(p,q)$ to 3), we do not expect the KS-statistic to be exactly zero, but they should also not be large.  In our experience with entropy balancing for continuous exposures, KS-statistics less than 0.1 are common and typically imply small differences in the tails of the distributions (in cases where the mean and variance have been matched).  If any of these values (correlation balance or marginal balance) appear large, then an optimal solution may not be possible and/or the data may not be appropriate for estimating a population dose-response curve.

Finally, we perform an assessment of the effective sample size as employing weights in an analysis typically reduces the sample information relative to an equal weighting; specifically greater variability in the weights corresponds to a greater loss of information. One reason for only focusing on balancing two or three moments is that by adding more moments we place more constraints on the optimization procedure. This often requires increasing the variability in weights and the effect is that this can make estimates of the dose-response relationship more variable.  The reduction in sample information is typically measured using the ``effective sample size (ESS)'' \citep{kish1965}, a measure from the survey literature related to the concept of a ``design effect''. The effective sample size is calculated as $ESS = \left(\sum_{i=1}^N w_i\right)^2 /\left({\sum_{i=1}^N w_i^2}\right)$ and in a binary setting would typically be calculated within each group, but in the continuous setting it is calculated using all weights. Within an applied analysis, the effective sample size can be thought of as the number of ``usable'' units that meet the constraints placed on the sample through weighting, i.e., if a unit's weight is effectively zero relative to all other unit's weights then that observation does not contribute to the analysis.  Additionally, checking the loss of information is useful for assessing the impact of weighting and determining if the available data support the analysis goals, i.e., small effective sample sizes between 0 and 100 imply low power and high variability in point estimates (i.e., low power will inflate the standard errors of the estimator).

\subsection{Estimation of Population Dose-Response Curves} \label{sec:locallinear}

The weights obtained in the previous section can be utilized in the same fashion as survey weights in order to estimate the population average dose-response curve; therefore we focus on flexible estimation techniques that can incorporate weights.  An overview of the estimation procedure in this paper is as follows: we utilize local linear regression \citep{cleveland1988,friedman2001elements} that is further weighted by entropy balancing weights to understand the population average behavior in a neighborhood around an exposure $A=a_0$ of interest.  We can then extend this to estimate the dose-response curve in the high-density region of the distribution of $A$.  We stress that our focus on flexible estimation of the balancing weights and the dose-response imply that we will only be able to estimate within the high density region of the exposure distribution and extrapolation past this high density region is not recommended.  

To estimate locally the method requires defining a target point of interest $A=a_0$ and selecting a kernel function $\kappa(a_0, \cdot)$ that defines the ``closeness'' between the target point $a_0$ and any other observed point $A_i = a$. Within this local region the model assumes a linear model where, 
\begin{align*}
    Y(a_0) \approx \beta_0(a_0) - \beta_1(a_0) A. 
\end{align*}
The coefficients $\beta_0(a_0)$ and $\beta_1(a_0)$ are therefore a function of the target point, i.e., they are locally adaptive.  Estimation then follows by minimizing a loss function within the neighborhood defined by the kernel function and further weighted by our entropy balancing weights, e.g., if $Y$ is continuous and choosing to minimize squared error loss,
\begin{align*}
    \min_{\beta_0(a_0), \beta_1(a_0)} \sum_{i=1}^N w_i \kappa(a_0, A_i) [Y_i - \beta_0(a_0) - \beta_1(a_0)A_i]^2
\end{align*}

The above requires specifying a kernel function $\kappa(a_0, \cdot)$.  A common example for local linear regression is the tricube function, $\kappa_{tc}(a_0, \cdot) = \left(1 - |u|^3\right)^3$, where $|u| \le 1$ and each $u_i$ is a normalized distance function, i.e., $u_i = d(a_0, A_i)/ \max_{j} d(a_0, A_j)$. The distance function $d(\cdot, \cdot)$ typically contains a nuisance parameter $\alpha$ that controls the fraction of points contained in the neighborhood of points used in estimating the outcome \citep{cleveland1988}. Other examples of kernels include the squared exponential or cosine similarity, but are not considered here \citep{scholkopf2001learning}.   It is possible to use cross-validation to select the parameter $\alpha$, or another model selection procedure, and in our application we use two-fold cross validation as each fold requires a large number of observations to minimize the sensitivity of the out-of-sample estimates across folds to the weight distribution.  If necessary, this procedure can be repeated multiple times to assess the distribution of the resulting $\alpha$ parameter.  

To implement these methods, we use the \texttt{loess} package in \texttt{R} that implements the tricube kernel and efficient estimation.  For an extensive overview of local regression methods in the context of prediction models see Chapter 6 in \cite{friedman2001elements}.  

\subsection{Inference using Bootstrapped Confidence Intervals} \label{sec:bootstrap}

Section \ref{sec:locallinear} provides point estimates of the average treatment effect in the population at specific points $A = a_0$, but does not provide adequate information for quantifying uncertainty.  Therefore to provide uncertainty estimates, we use a bootstrap procedure \citep{efron1994introduction} focused on bootstrapping the entire estimation procedure.  Specifically for $b = 1, \dots, B$, we sample with replacement from the observations to create a new data set $\mathcal{D}^b = \{Y^b, A^b, \bs{X}^b\}$ and using this new sample to re-estimate the entropy balancing weights and perform the entire dose-response estimation procedure on the outcomes $Y^b$ to obtain a new estimate $\hat f^b(A_i)$ for all $i$.  That is, we perform cross-validation to select a new $\alpha_b$, and refit the local linear model to obtain a new estimate of the treatment effect at that point.  This process is repeated for the set of $B$ bootstrap samples and the standard error of the bootstrap distribution of estimates at the point $A=a_0$, denoted $se(\hat f^b(a_0))$, is calculated. Approximate confidence intervals can then be constructed, i.e., 
\begin{align}
    [\mathcal{L}(A), \mathcal{U}(A)] = \left[\hat f(A) - 2 \times se(\hat f^b(A)), \hat f(A) + 2 \times se(\hat f^b(A)) \right] \label{eqn:bootstrapinterval}
\end{align}
We evaluate properties of this interval estimator in Section \ref{sec:coverage}.

\section{Simulation} \label{sec:simulation}

\subsection{Data Generation}

To demonstrate the method we perform a simulation study that attempts to stress the method in reasonable ways, i.e., non-linear relationship with $A$ in the outcome model and a strictly positive treatment variable. We generate 1000 simulations where for each replication we generate a sample size of $N=1000$.  

We generate data with three covariates, 
\begin{align*}
    X_1 \sim \mathcal{N}(-0.5, 1)  && X_2 \sim \mathcal{N}(1, 1) && X_3 \sim \mbox{Bernoulli}(p_3 = 0.3) 
\end{align*}
and relate these to treatment in the following way,
\begin{align*}
    A \sim \chi^2(df = 3, \mu_A) && \mu_A \equiv 5 |X_1| + 6 |X_2| + 3 |X_3| 
\end{align*}
where $\mu_A$ is a non-centrality parameter that creates a somewhat realistic treatment distribution generated from a mixture of covariate types, i.e., high concentration of treatment at zero with a longer right tail of higher valued treatments, where an example of such an exposure may be the cumulative number of procedures that a client is exposed to during a substance abuse treatment program similar to our motivating example.  

The outcome is generated as follows,
\begin{align*}
    Y &= - \frac{1}{300} (A - 5)(A + 5) + \frac{1}{25} A (X_1^2 + X_2^2) + X_1 + X_2 + X_3 + \epsilon, \hspace{1em} \\
    \epsilon &\sim \mathcal{N}(0,1)
\end{align*}
which implies that the marginal dose-response curve is 
\begin{align*}
    E[Y(a)] = - \frac{1}{300} (A - 5)(A + 5) + \frac{13}{100} A + 0.8 
\end{align*}
This outcome model introduces non-linearity in $A$ that would be difficult to capture with traditional modeling and includes an interaction between treatment and covariates that can cause a strong bias when only modeling the marginal relationship between $A$ and $Y$ and there is confounding by $\bs{X}$. An additional simulation is included in the appendix that considers the case where there is no effect of treatment.  

\begin{figure}[h]
    \centering
    \includegraphics[width = \textwidth]{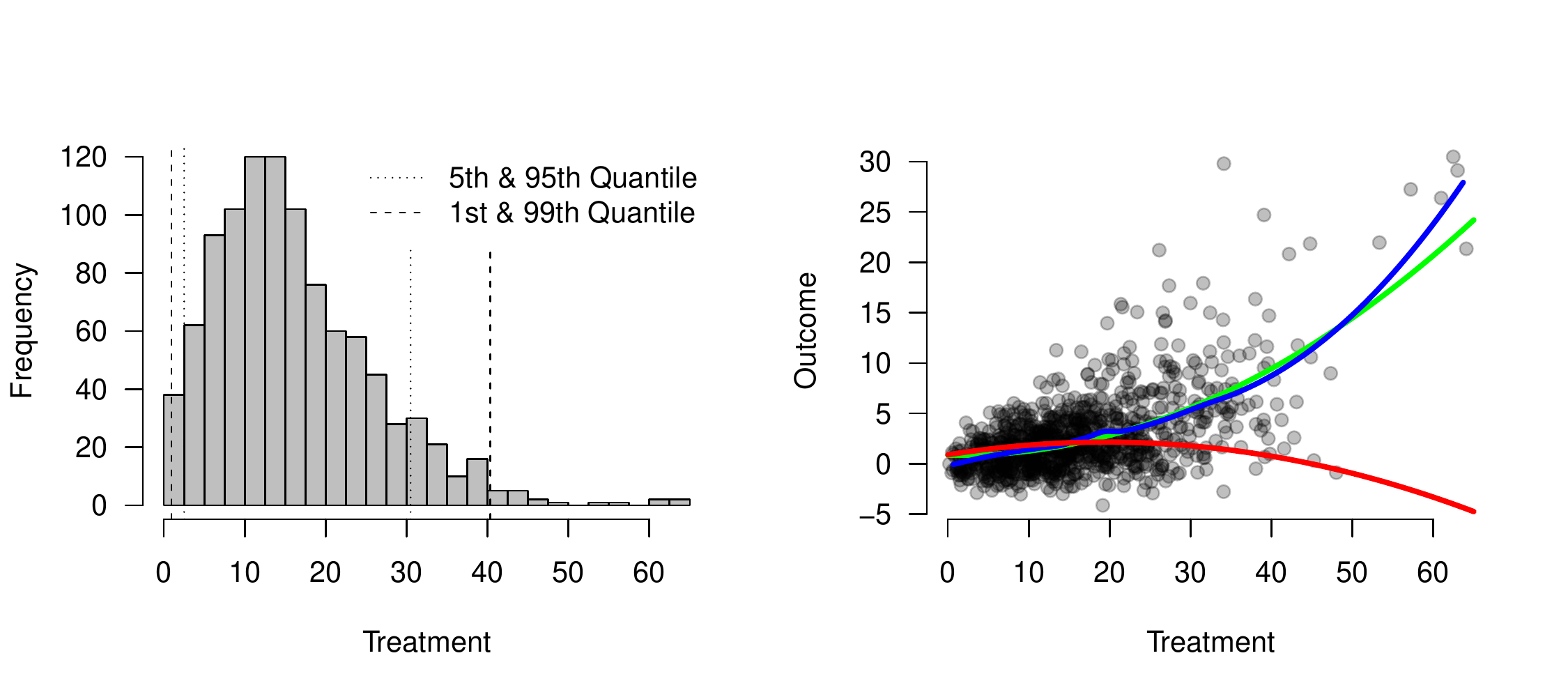}
    \caption{Visualization of the distribution of exposure variable (left) and relationship between exposure and outcome (right).  In the left figure, the high-density region of $A$ is highlighted.  In the right figure, the true marginal relationship is shown in {\color{red} \textbf{red}}, a simple unweighted linear smoother is shown in {\color{blue} \textbf{blue}}, and an unweighted quadratic estimation is provided in {\color{green} \textbf{green}}.}  
    \label{fig:explore1}
\end{figure}

The left panel of Figure \ref{fig:explore1} demonstrates the marginal distribution of $A$ and the right panel demonstrates the observed relationships between $A$ and $Y$ with the true population marginal dose-response curve overlain and a linear smoother through the data.  The figure demonstrates that there is confounding for this population dose-response curve, where if $A$ and $\bs{X}$ did not have strong relationships, then the curves would be similar. Additionally, the trend in the unweighted distribution is in the opposite direction, i.e., in the observed relationship there appears to be a positive association between exposure and outcome when the true marginal relation (averaged over the population) is a negative quadratic association.  

\subsection{Estimation Methods Compared}

To fairly evaluate the method, we compare entropy balancing against other methods of estimating weights; these are listed in Table \ref{tab:altweights}.  We primarily focus on comparing against two ``naïve'' type analyses and an alternative using the generalized Covariate Balancing Propensity Score (CBPS) of \cite{fong2018covariate} and implemented in the \texttt{CBPS} \texttt{R} package; specifically their fully nonparametric version.  For both entropy balancing and CBPS, we perform the estimation focusing on a range of moments balanced.  

\begin{table}[ht!]
    \centering
    \begin{tabular}{c|l}
         & Weight Estimation Method \\ \hline 
        1) & Unweighted estimation; i.e., naïve analysis  \\
        2) & Entropy balancing (1); as described above, one moment  \\ 
        & -One moment of $\bs{X}$ covariance balanced  (i.e., $p=1$) with $A$ ($q=1$) \\
        & -One marginal moment balanced (i.e., $p,q=1$) \\ 
        3) & Entropy balancing (2); two moments.  \\ 
        & -Two moments of $\bs{X}$ covariance balanced  (i.e., $p=1,2$) with $A$ ($q=1$) \\
        & -Two marginal moments balanced (i.e., $p,q=1,2$) \\ 
        4) & Entropy balancing (3); three moments.  \\ 
        5) & Entropy balancing (4); four moments.  \\ 
        6) & Parametric assumptions: $p(a|\bs{x})$ is a linear model with first-order terms,\\
        & and normally distributed errors; $p(a)$ is normally distributed. \\ 
        7) & Generalized CBPS (1): nonparametric assumptions, cor. prior set to 1e-8, one moment \\ 
        & Model definition:  \texttt{A $\sim$ X1 + X2 + X3} \\ 
        8) & Generalized CBPS (2): nonparametric assumptions, cor. prior set to 1e-8, two moments \\ 
        & Model definition:  \texttt{A $\sim$ poly(X1,2) + poly(X2,2) + X3} \\ 
        9) & Generalized CBPS (3): nonparametric assumptions, cor. prior set to 1e-8, three moments \\ 
        10) & Generalized CBPS (4): nonparametric assumptions, cor. prior set to 1e-8, four moments \\ 
    \end{tabular}
    \caption{Descriptions of the weight estimation procedures.}
    \label{tab:altweights}
\end{table}

To evaluate the performance of the dose-response curve, we propose estimation using the suggested local linear regression method, as well as a parametric ``global'' linear model assumption where $E[Y|A] = \beta_0 + \beta_1 A + \beta_2 A^2$ to compare relative performance between the nonparametric methods and a model with the correct form of the marginal relationship, i.e., to demonstrate performance where the only correct specification required is for the weights.  For each outcome estimation we use the weights as described above.  For a fair evaluation across methods, we perform a two-fold cross validation to select $\alpha$ for each method. Finally, to compare estimates across the methods, we only compare predictions within the approximate high-density region of $A$, i.e., $A=a_0\in [0,45]$, shown in Figure \ref{fig:explore1}.  

\subsection{Results} 

\subsubsection{Performance Balancing Covariates}

The first result compares correlation balance, specifically the correlation between each covariate and the exposure variable.  This is illustrated in Table \ref{tab:covariatebalance} where the correlation balance statistics for the entropy balancing method are all less than $1e-3$ (both the average across replications and the maximum value attained).  We also see that the method provides good performance in retaining the marginal distributions under weighting. The KS-statistics across replications are all less than 0.1, which is often considered a good guideline for adequate covariate balance in binary settings.  Table \ref{tab:covariatebalance} also demonstrates that using misspecified linear models can provide unpredictable behavior and erratic covariate imbalance measures, where the average absolute weighted correlation is higher than what was observed in the sample.  Additionally, there are large discrepancies between the weighted covariate distributions and the observed distributions.  Finally, the CBPS methods perform similar to the entropy balancing method.  The CBPS methods obtain higher weighted correlations on average than entropy balancing, but the values would be very acceptable in practice.  

\begin{table}[ht]
\centering
\resizebox{\linewidth}{!}{
\begin{tabular}{l|ccc|ccc|cccc}
    & \multicolumn{3}{c|}{Conditional Balance}  & \multicolumn{3}{c|}{Correlation Balance} & \multicolumn{4}{c}{Marginal Balance: KS statistics only} \\
  \hline\hline 
  & \multicolumn{7}{c}{\textbf{Average Across Replications}} \\  
  Method & $X_1$ & $X_2$ & $X_3$ & $X_1$ & $X_2$ & $X_3$ & $A$ & $X_1$ & $X_2$ & $X_3$ \\ 
  \hline
Unweighted & 0.021 & 0.045 & 0.007 & 0.201 & 0.428 & 0.142 & 0.000 & 0.000 & 0.000 & 0.000 \\ 
  Entropy Balancing (1) & 0.000 & 0.000 & 0.000 & 0.000 & 0.000 & 0.000 & 0.047 & 0.019 & 0.038 & 0.000 \\ 
  Entropy Balancing (2) & 0.000 & 0.000 & 0.000 & 0.000 & 0.000 & 0.000 & 0.030 & 0.025 & 0.037 & 0.000 \\ 
  Entropy Balancing (3) & 0.000 & 0.000 & 0.000 & 0.000 & 0.000 & 0.000 & 0.026 & 0.023 & 0.025 & 0.000 \\ 
  Entropy Balancing (4) & 0.000 & 0.000 & 0.000 & 0.000 & 0.000 & 0.000 & 0.020 & 0.022 & 0.025 & 0.000 \\ 
  Linear Model & 0.025 & 0.040 & 0.006 & 0.257 & 0.346 & 0.184 & 0.245 & 0.205 & 0.268 & 0.079 \\ 
  CBPS - Nonparametric: (1) & 0.000 & 0.000 & 0.000 & 0.003 & 0.002 & 0.002 & 0.026 & 0.027 & 0.035 & 0.001 \\ 
  CBPS - Nonparametric: (2) & 0.000 & 0.000 & 0.000 & 0.003 & 0.004 & 0.003 & 0.044 & 0.032 & 0.032 & 0.001 \\ 
  CBPS - Nonparametric: (3) & 0.000 & 0.001 & 0.000 & 0.004 & 0.004 & 0.004 & 0.046 & 0.032 & 0.031 & 0.002 \\ 
  CBPS - Nonparametric: (4) & 0.001 & 0.001 & 0.001 & 0.009 & 0.010 & 0.010 & 0.047 & 0.031 & 0.031 & 0.004 \\ 
   \hline\hline
  & \multicolumn{7}{c}{\textbf{Maximum Across Replications}} \\
   Method & $X_1$ & $X_2$ & $X_3$ & $X_1$ & $X_2$ & $X_3$ & $A$ & $X_1$ & $X_2$ & $X_3$ \\ 
  \hline
Unweighted & 0.034 & 0.054 & 0.011 & 0.324 & 0.517 & 0.242 & 0.000 & 0.000 & 0.000 & 0.000 \\ 
  Entropy Balancing (1) & 0.000 & 0.000 & 0.000 & 0.000 & 0.000 & 0.000 & 0.077 & 0.035 & 0.055 & 0.000 \\ 
  Entropy Balancing (2) & 0.000 & 0.000 & 0.000 & 0.000 & 0.000 & 0.000 & 0.069 & 0.052 & 0.064 & 0.000 \\ 
  Entropy Balancing (3) & 0.000 & 0.000 & 0.000 & 0.000 & 0.000 & 0.000 & 0.049 & 0.046 & 0.054 & 0.000 \\ 
  Entropy Balancing (4) & 0.000 & 0.000 & 0.000 & 0.001 & 0.002 & 0.000 & 0.044 & 0.048 & 0.076 & 0.000 \\ 
  Linear Model & 0.153 & 0.192 & 0.031 & 0.919 & 0.930 & 0.824 & 0.995 & 0.986 & 0.995 & 0.698 \\ 
  CBPS - Nonparametric: (1) & 0.005 & 0.018 & 0.002 & 0.043 & 0.145 & 0.043 & 0.062 & 0.052 & 0.074 & 0.014 \\ 
  CBPS - Nonparametric: (2) & 0.004 & 0.006 & 0.001 & 0.040 & 0.058 & 0.028 & 0.090 & 0.066 & 0.056 & 0.011 \\ 
  CBPS - Nonparametric: (3) & 0.006 & 0.007 & 0.002 & 0.051 & 0.064 & 0.053 & 0.095 & 0.069 & 0.061 & 0.012 \\ 
  CBPS - Nonparametric: (4) & 0.011 & 0.014 & 0.007 & 0.091 & 0.124 & 0.131 & 0.093 & 0.095 & 0.080 & 0.061 \\ 
\end{tabular}
}
\caption{Summaries of balance assessments across replications. Conditional balance is assessed here using the absolute value of the coefficient regressing $A$ on $X_j$ in a linear model.  Covariance balance is assessed using the absolute value of the weighted correlation between each variable and the exposure variable $A$.  The marginal balance is assessed using the Kolmogorov–Smirnov statistic between the unweighted empirical CDF and the weighted CDF. The top portion of the table are the averages of these statistics across replications and the bottom portion contain the maximum across replications. }
\label{tab:covariatebalance}
\end{table}

\subsubsection{Performance in Estimating Dose-Response Curves} \label{sec:curveests}

\begin{figure}[ht!]
    \centering
    \includegraphics[width=\textwidth]{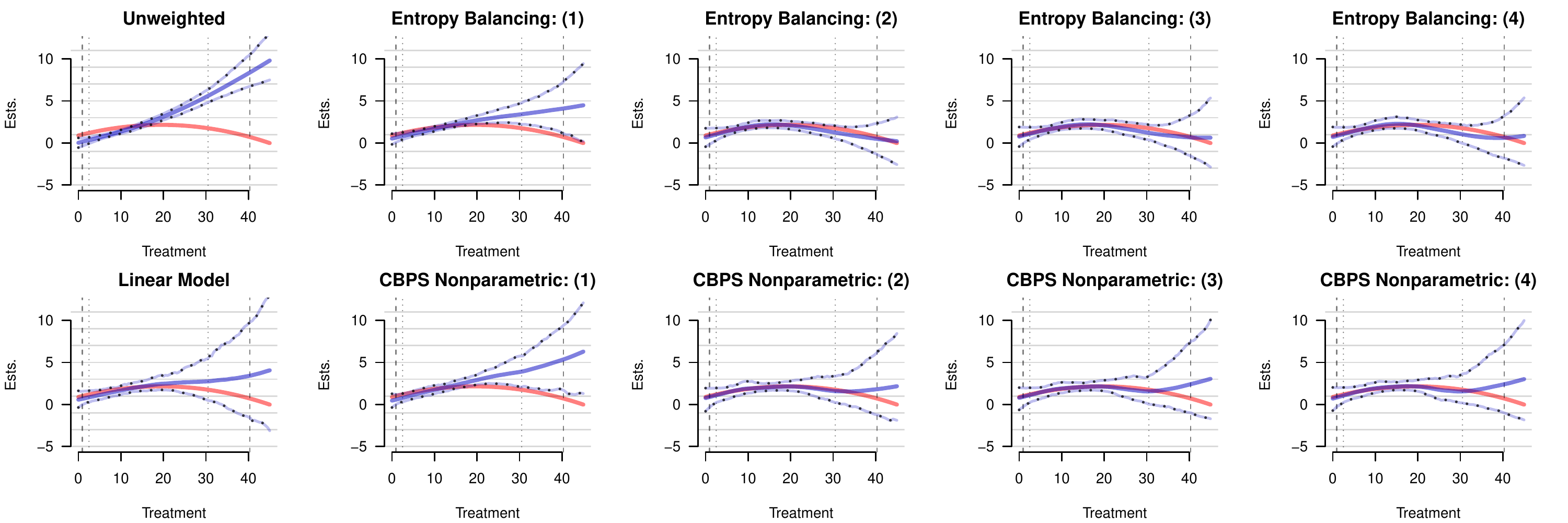}
    \caption{Performance estimating the dose-response curve across repeated simulated samples: Local Linear Regression. The {\color{red} \textbf{red}} line is the true population dose-response curve. The {\color{blue} \textbf{blue}} lines represent the estimated curves; the solid line is the mean across replications and the dotted lines represent a 95\% equal-tail interval of the density of estimates. }
    \label{fig:response1}
\end{figure}

\begin{figure}[ht!]
    \centering
    \includegraphics[width=\textwidth]{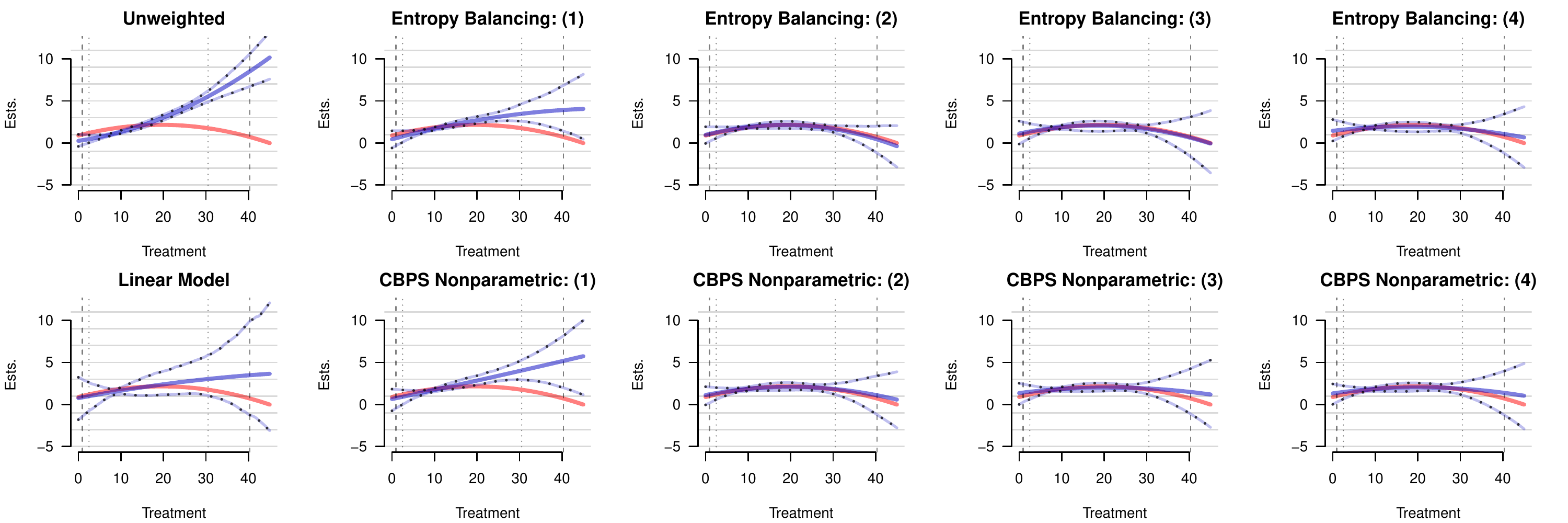}
    \caption{Performance estimating the dose-response curve across repeated simulated samples: Linear Regression - second order polynomial in treatment. The {\color{red} \textbf{red}} line is the true population dose-response curve. The {\color{blue} \textbf{blue}} lines represent the estimated curves; the solid line is the mean across replications and the dotted lines represent a 95\% equal-tail interval of the density of estimates. }
    \label{fig:response2}
\end{figure}

While assessing covariate balance is an important step in a causal inference analysis, the weights are often nuisance parameters in the estimation of the effect of interest, i.e., in this application the population dose-response curve.  Figures \ref{fig:response1} and \ref{fig:response2} illustrate visually the performance of the methods in estimating the dose-response curves, i.e., the sampling distribution across the simulation replications.  In these figures, it is clear that the unweighted estimates provide curves that would incorrectly indicate a positive association despite the true marginal relationship being a negative quadratic association. Methods that focus on balancing a higher number of moments (i.e., 2 or 3) appear to capture the true relationship well in this simulation and the average of the curves across the simulations within the high-density region match the true marginal relationship.  Additionally, the distribution across the curves demonstrates that there is not a high degree of variability in these estimates despite the effective sample size being reduced to a small fraction of the original number of observations on average.  For methods that focus on only balancing one moment, we see that these weights reduce the bias in estimating the population dose-response curve, but that they do not capture the true relationship for high exposure values, i.e., $A > 20$.  

\begin{table}[h!]
    \centering 
    \begin{tabular}{l|c|c|c|c|c}
    && \multicolumn{2}{c|}{Average Bias} & \multicolumn{2}{c}{Mean Square Error}\\\hline
    Weighting Method & Avg. ESS & LOESS & Reg. & LOESS & Reg. \\\hline 
  Unweighted & 1000.000 & 2.564 & 2.568 & 17.415 & 17.817 \\ \hdashline
  Entropy Balancing (1) & 737.228 & 1.166 & 1.133 & 4.034 & 3.637 \\ 
  Entropy Balancing (2) & 463.829 & -0.172 & -0.078 & 0.361 & 0.234 \\ 
  Entropy Balancing (3) & 406.130 & -0.128 & -0.037 & 0.573 & 0.459 \\ 
  Entropy Balancing (4) & 385.312 & -0.172 & 0.065 & 0.703 & 0.496 \\ \hdashline
  Linear Model & 137.388 & 0.830 & 0.886 & 4.738 & 5.468 \\ \hdashline 
  CBPS - Nonparametric: (1) & 441.807 & 1.624 & 1.594 & 7.406 & 6.573 \\ 
  CBPS - Nonparametric: (2) & 243.853 & 0.212 & 0.112 & 1.334 & 0.420 \\ 
  CBPS - Nonparametric: (3) & 219.252 & 0.340 & 0.212 & 2.141 & 0.710 \\ 
  CBPS - Nonparametric: (4) & 225.603 & 0.334 & 0.186 & 2.157 & 0.648 \\ 
    \end{tabular}
    \caption{Comparisons of the average bias and mean squared error between the estimated dose-response curve and the true marginal relationships on a grid $a_0 \in [0,45]$ and the average of the effective samples sizes for each method.   Within the bias and mean squared error sections of the table, the left column demonstrates the proposed local linear regression method and the right column are results using a linear regression if the correct model for the marginal relationship were known.}
    \label{tab:results1}
\end{table}

Table \ref{tab:results1} contains the effective sample size in estimation and comparisons of the average bias and mean squared error between the true marginal dose-response curve on a grid of points in $A\in[0,45]$ and the estimated curves across the simulations.  These results largely mimic the visual argument of the figures, but provide a quantification of the differences.  Estimation using local linear regression and balancing on more than one moment provides better performance than balancing only one moment; this finding is observed for both the proposed entropy balancing method and the CBPS.  In this example, the entropy balancing method with higher moments generally performs better than the CBPS method when using local linear regression, but both methods are more similar when the true outcome model is known.  One surprise is that despite the poor covariate balance, the linear model performed better than CBPS when only one moment is matched in this simulation setting.  All methods reduce the MSE relative to an unweighted analysis because they are controlling for confounding variables through weighting.

\subsection{Coverage of Bootstrap Confidence Intervals} \label{sec:coverage}

The results of the previous two sections demonstrate that the entropy balancing procedure is both able to provide adequate covariate balance and in some instances better performance in providing point estimates for the population dose-response curve in the high density region of the exposure. A point estimate though is often only part of the inferential story and a rigorous analysis requires a quantification of the uncertainty about that point estimate.  To that end, we construct a simulation study that examines the performance of the bootstrap procedure defined in Section \ref{sec:bootstrap}.  

\begin{figure}
    \centering
    \includegraphics[width=\linewidth]{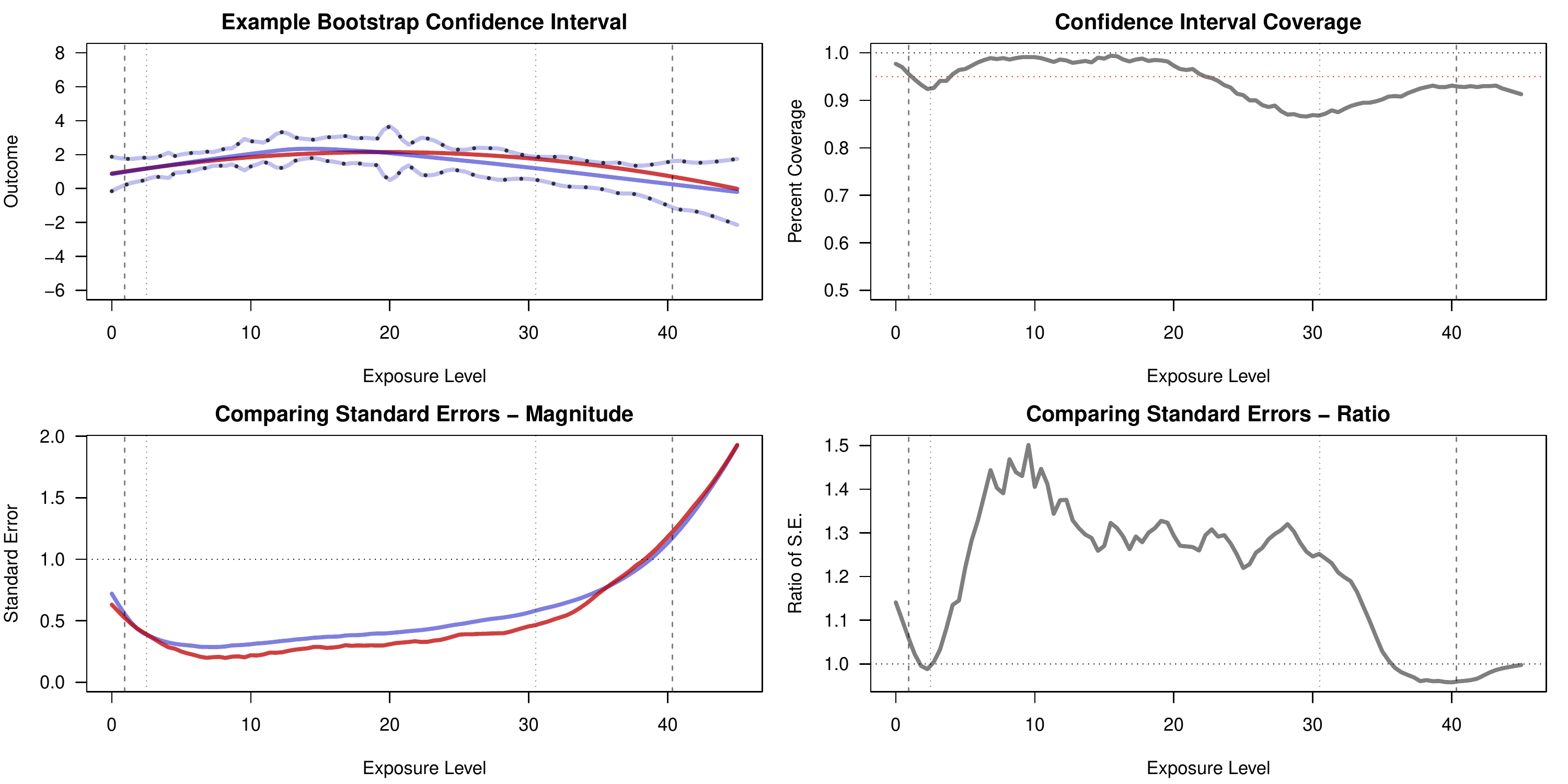}
    \caption{Simulation results for bootstrap confidence intervals. Upper left panel contains a single visualization of a bootstrapped confidence interval for the curve. Throughout the {\color{red} \textbf{red}} curves represent the ``truth'' and the  {\color{blue} \textbf{blue}} solid line represents the estimated curve; the dotted blue lines represent estimated 95\% confidence intervals.  Upper right panel contains the point-wise coverage of the 95\% bootstrapped confidence intervals.  Lower panels contain the magnitude (left) and ratio (right) of the average point-wise bootstrap standard error across the bootstrap simulations as compared to the standard error of the estimated curves obtained in this simulation. In all figures the vertical lines represent the $1^{st}, 5^{th}, 95^{th}, 99^{th}$ quantiles of the distribution of the exposure variable $A$ in the high-density region.}
    \label{fig:bootstrap1}
\end{figure}

In this simulation we use the data-generation procedure outlined in Section \ref{sec:simulation} and again construct 1000 replications, but for each estimation procedure also provide bootstrap confidence intervals using $B=100$ replications.  The results from this study are provided in Figure \ref{fig:bootstrap1}. The upper left panel provides a visualization of a single bootstrap confidence interval for the provided points. The upper right panel provides the coverage of the 95\% confidence intervals constructed using the bootstrap standard errors and our approximate intervals defined in Equation \ref{eqn:bootstrapinterval}. The lower left panel compares the average bootstrap standard error obtained with the proposed procedure to the standard error of the estimates across the 1000 corresponding estimates of the curve obtained in the simulation. The lower right panel contains the ratio of the average bootstrap standard error obtained with the proposed procedure to the standard error of the estimates.  

The results demonstrate that the standard errors from the bootstrap estimation procedure are slightly conservative when compared with the sampling distribution of the estimates (lower right panel).  This results in confidence intervals that are often too wide and therefore provide greater than $95\%$ coverage as demonstrated in the right panel in the high density region of the exposure. Coverage suffers towards the tails where there may be some bias in estimation due to sparsity of observations in this region. Future work should examine the properties of this procedure for estimating uncertainty in weighted population dose-response curves.  

\section{Application} \label{sec:application}

\subsection{Overview of the case study data and analysis}

We provide a demonstration of these methods using data collected from four cohorts of the Assertive Adolescent and Family Treatment Initiative (AAFTI) that had, as part of its initiative, a goal of assessing the effect of the Adolescent Community Reinforcement Approach (A-CRA) \citep{godley2016adolescent,godley2011largescale} on longitudinal substance use and behavioral outcomes among adolescent substance users. A-CRA is a community-based reinforcement behavioral treatment program for alcohol and substance use disorders administered over a three-month period that recommends 10 individual focused sessions between a clinician and a client, two sessions between the clinician and at least one caregiver, and an additional two sessions that include all parties. However, clinicians are able to suggest as many sessions as necessary within an individual treatment program.  Within each session the therapist performs a set of procedures from a therapy guide (outlined in \cite{godley2016adolescent}), such as the ``Problem Solving Skills Training'' designed to improve the clients ability to solve problems outside of therapy, the ``Happiness Scale'' used to understand levels of happiness in various aspects of the clients life, and ``Increasing Prosocial Recreation'' that focuses on developing behaviors that increase interactions with individuals that are less likely to lead to drug use, among many other procedures.  While the therapy guide is a suggested course of treatment, each trained clinician administering A-CRA is able to use additional sessions if necessary and has latitude in how they choose the many procedures at their disposal during treatment sessions. Individuals within AAFTI were then also exposed to Assertive Continuing Care, that had A-CRA embedded within its sessions, for an additional three months; providing a total of sixth months of A-CRA exposure. Among those adolescents engaging in treatment, investigating the relationship between the number of sessions an individual receives and emotional and substance use outcomes is of interest. In the AAFTI, clinicians engaged in 4 to 79 sessions, where 4 sessions was a criteria for engagement with the substance use treatment.  Individuals that were identified as ``not engaging'' with treatment were removed from the analysis.  

Throughout the implementation of A-CRA, covariate information on the client is collected and additional targeted information related to psychological risk factors and substance use are collected through the Global Appraisal of Individuals Needs (GAIN) \citep{dennis2003global}, a survey composed of measures that attempt to capture domains such as substance use, physical health, mental and emotional health, living situation, and legal situation. Within AAFTI, both covariate information and the GAIN were measured at baseline, 3 months, 6 months, and 12 months, providing longitudinal assessment over a one-year period.  

\begin{figure}[H]
    \centering
    \includegraphics[width=\textwidth]{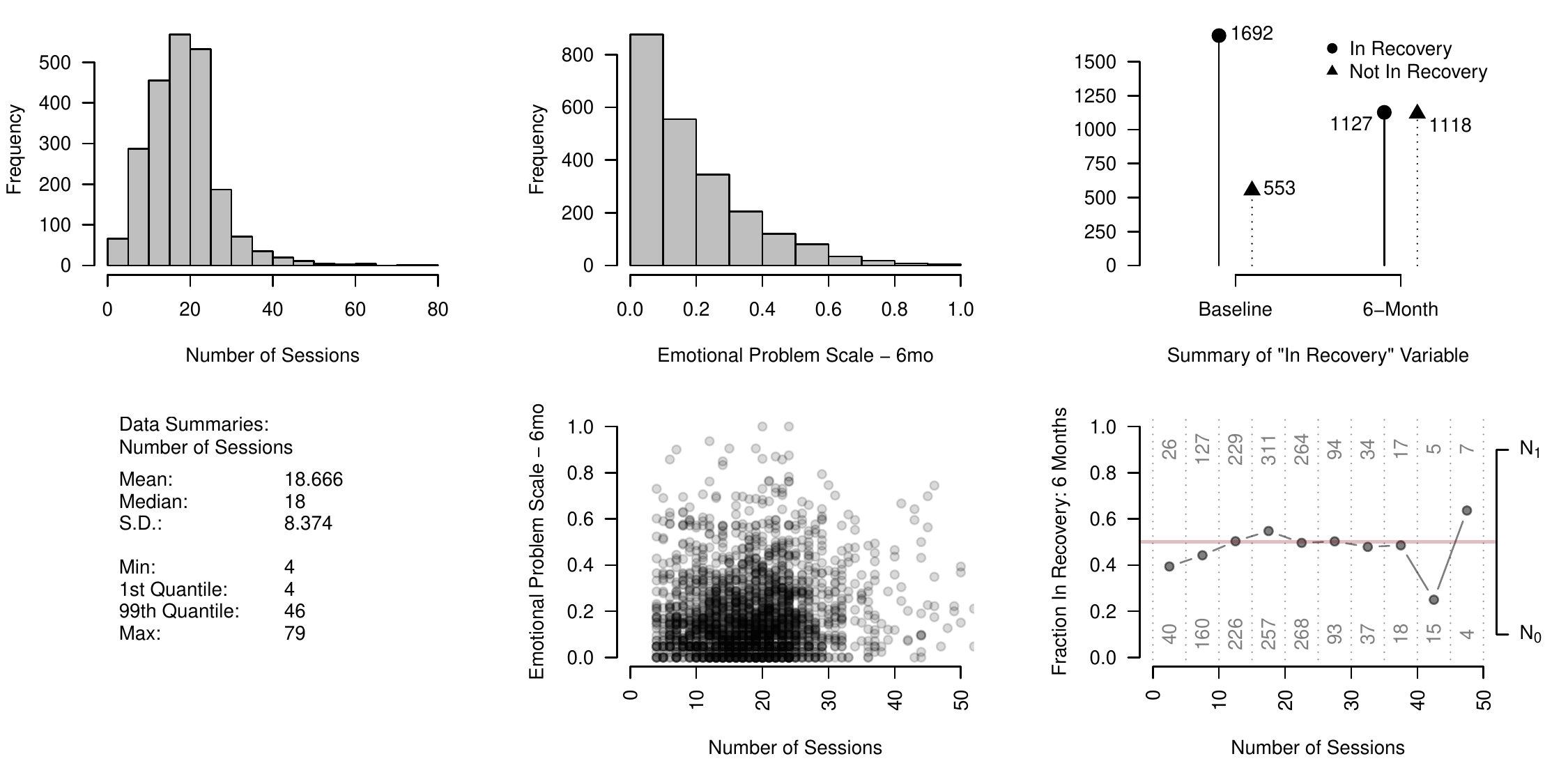}
    \caption{Relationships among treatment exposure (number of sessions) and outcomes of interest.  The lower figures are focused only on the highest density region of the treatment exposure, i.e., number of sessions $ \in [4,50]$ }
    \label{fig:app1}
\end{figure}

In this analysis we focus on assessing the marginal relationship between the number of treatment sessions delivered to the adolescent client on two outcome variables of interest at the 6-month follow-up point: 1) the emotional problems scale (EPS) of the GAIN, and 2) the binary indicator of whether the individual is ``in recovery'' or not. The variable ``in recovery'' is defined as being one (true) if the client has been housed in the community, has been abstinent from any substance use within the past month, and reports no past month substance use problems.  Figure \ref{fig:app1} demonstrates that there is an increase of those in recovery at six months when compared to baseline.  

In addition to the primary exposure (i.e., number of treatment sessions) and outcome variables, we consider baseline covariate information such as age, gender, race, and a subset of the GAIN scales that have been identified as being necessary to control for in our analysis. Figure \ref{fig:app1} contains visualizations of the initial distributions of key variables (exposure and outcomes) to highlight similarities with our simulation study. One feature of the exposure distribution is that the 99th quantile is approximately 46, so we do not expect to trust inference from our method much higher than this value, as we would be extrapolating to a part of the exposure distribution with low representation. We estimate the dose-response curve in the range [4,50] and remove individuals not in this range when performing the analysis.  Within the original data set, there is missingness within some of the covariates. In this analysis, we analyse a single imputed data set from a larger imputation procedure of these data used to test the relative effectiveness of A-CRA versus another evidence based treatment program for adolescent substance users \citep{griffin2020}. As we are focused on highlighting and demonstrating the utility of the method, we leave discussions of the full implementation of these methods with missing data to future work. Table \ref{tab:apptab} contains a detailed list of the covariate information and baseline GAIN measures that were included in this analysis.    

There are two complications related to our primary exposure variable that complicate an applied causal analysis in our case study: 1) the exposure variable is not decided at entry, that is, there are factors related to the number of treatment sessions that cannot be observed longitudinally throughout exposure to A-CRA, such as client and clinician interactions, and 2) there can be multiple versions of the same treatment level as we do not know the combination of the underlying procedures used (e.g., two clients each exposed to 10 sessions could have an underlying different set of procedures that they were exposed to).  These issues introduce complications in interpreting our analysis as causal that we discuss further Section 4.3. Regardless, the entropy balancing weights should provide a robust associational relationship that marginalizes over, i.e., controls for, the observed covariates and will yield useful insights for clinicians about the relationship between total number of A-CRA treatment sessions and our outcomes. 

\begin{table}[ht!]
    \centering
\resizebox{\linewidth}{!}{
\begin{tabular}{l|cc|cccc|cc|cccc|c}
& \multicolumn{6}{c|}{Unweighted} & \multicolumn{6}{c|}{Weighted} & \\
  \hline
 \textbf{Variable} & Mean & SD & Cor. & $\beta$ & $t$ & $p$ & Mean & SD & Cor. & $\beta$ & $t$ & $p$ & KS \\ 
  \hline
  \textbf{Exposure Variable}: Number of Sessions & 18.38 & 7.65 & - & - & - & - & 18.38 & 7.65 & - & - & - & - & 0.00 \\ \hline
  \textbf{Demographic Variables} &&&&&&&&&&&&& \\ 
  Gender ($1\equiv$ Male) & 0.73 & 0.44 & -0.09 & -0.04 & -4.12 & 0.00 & 0.73 & 0.44 & 0.00 & 0.00 & 0.00 & 1.00 & 0.00 \\
  Age & 15.54 & 1.25 & -0.08 & -0.10 & -3.64 & 0.00 & 15.54 & 1.25 & 0.00 & 0.00 & 0.00 & 1.00 & 0.00 \\ 
  Race &&&&&&&&&&&&& \\ 
  \hspace{0.5em} \textit{White/Caucasian} & 0.34 & 0.47 & 0.06 & 0.03 & 2.76 & 0.01 & 0.34 & 0.47 & 0.00 & 0.00 & 0.00 & 1.00 & 0.00 \\ 
  \hspace{0.5em} \textit{Black/African-American} & 0.12 & 0.33 & -0.03 & -0.01 & -1.56 & 0.12 & 0.12 & 0.33 & -0.00 & -0.00 & -0.00 & 1.00 & 0.00 \\ 
  \hspace{0.5em} \textit{Hispanic} & 0.37 & 0.48 & -0.03 & -0.01 & -1.34 & 0.18 & 0.37 & 0.48 & -0.00 & -0.00 & -0.00 & 1.00 & 0.00 \\ 
  \hspace{0.5em} \textit{Other} & 0.17 & 0.38 & -0.01 & -0.00 & -0.40 & 0.69 & 0.17 & 0.38 & 0.00 & 0.00 & 0.00 & 1.00 & 0.00 \\ 
  \hline
  \textbf{Baseline GAIN Variables} &&&&&&&&&&&&& \\
  Ever been victimized? & 0.58 & 0.49 & 0.03 & 0.01 & 1.20 & 0.23 & 0.58 & 0.49 & 0.00 & 0.00 & 0.00 & 1.00 & 0.00 \\
  Substance Frequency Scale & 12.00 & 13.65 & -0.01 & -0.10 & -0.36 & 0.72 & 12.00 & 13.65 & 0.00 & 0.00 & 0.00 & 1.00 & 0.00 \\ 
  In Detoxification program or not (past 90 days) & 0.03 & 0.17 & 0.02 & 0.00 & 1.12 & 0.26 & 0.03 & 0.17 & 0.00 & 0.00 & 0.00 & 1.00 & 0.00 \\ 
  Current Withdrawal Scale-Past Week & 1.05 & 2.98 & 0.02 & 0.06 & 0.97 & 0.33 & 1.05 & 2.98 & 0.00 & 0.00 & 0.00 & 1.00 & 0.00 \\ 
  Needle Use & 0.07 & 0.60 & -0.03 & -0.02 & -1.40 & 0.16 & 0.07 & 0.60 & 0.00 & 0.00 & 0.00 & 1.00 & 0.00 \\ 
  Worsening Physical Health due to SU & 0.65 & 0.97 & 0.01 & 0.01 & 0.25 & 0.80 & 0.65 & 0.97 & 0.00 & 0.00 & 0.00 & 1.00 & 0.01 \\ 
  Cognitive Impairment Screen & 3.55 & 3.21 & -0.02 & -0.05 & -0.72 & 0.47 & 3.55 & 3.21 & -0.00 & -0.00 & -0.00 & 1.00 & 0.00 \\ 
  Depressive Symptom Scale & 2.75 & 2.64 & 0.06 & 0.15 & 2.63 & 0.01 & 2.74 & 2.64 & 0.00 & 0.00 & 0.00 & 1.00 & 0.00 \\ 
  Mental Health Treatment (past 90 days) & 0.26 & 0.44 & 0.08 & 0.04 & 3.86 & 0.00 & 0.26 & 0.44 & 0.00 & 0.00 & 0.00 & 1.00 & 0.00 \\ 
  Homicidal Suicidal Thought Scale & 0.28 & 0.84 & 0.00 & 0.00 & 0.20 & 0.84 & 0.28 & 0.84 & 0.00 & 0.00 & 0.00 & 1.00 & 0.00 \\ 
  General Conflict Tactic Scale & 2.99 & 2.78 & 0.01 & 0.03 & 0.50 & 0.62 & 2.99 & 2.78 & 0.00 & 0.00 & 0.00 & 1.00 & 0.00 \\ 
  Continued substance use despite prior tx & 0.09 & 0.29 & 0.00 & 0.00 & 0.04 & 0.97 & 0.09 & 0.29 & 0.00 & 0.00 & 0.00 & 1.00 & 0.00 \\ 
  Substance Abuse Tx Index & 5.14 & 17.67 & 0.02 & 0.32 & 0.86 & 0.39 & 5.14 & 17.66 & -0.00 & -0.00 & -0.00 & 1.00 & 0.00 \\ 
  Substance Use Dependence (past year) & 4.29 & 3.45 & 0.05 & 0.16 & 2.21 & 0.03 & 4.29 & 3.45 & 0.00 & 0.00 & 0.00 & 1.00 & 0.00 \\ 
  Living Environmental Risk Index & 10.73 & 2.80 & 0.04 & 0.12 & 2.07 & 0.04 & 10.73 & 2.80 & 0.00 & 0.00 & 0.00 & 1.00 & 0.00 \\ 
  Social Environmental Risk Index & 13.09 & 4.28 & 0.00 & 0.01 & 0.10 & 0.92 & 13.09 & 4.28 & 0.00 & 0.00 & 0.00 & 1.00 & 0.00 \\ 
  Recovery Environment Risk Index & 19.74 & 27.96 & 0.06 & 1.69 & 2.85 & 0.00 & 19.74 & 27.95 & 0.00 & 0.00 & 0.00 & 1.00 & 0.00 \\ 
  Parent Activity Index & 3.27 & 1.33 & 0.03 & 0.04 & 1.31 & 0.19 & 3.27 & 1.33 & -0.00 & -0.00 & -0.00 & 1.00 & 0.00 \\ 
  Adjusted Days Abstinent (past 90 days) & 50.15 & 32.78 & 0.00 & 0.16 & 0.23 & 0.82 & 50.15 & 32.77 & -0.00 & -0.00 & -0.00 & 1.00 & 0.00 \\ 
  Days in a controlled environment (past 90 days) & 10.23 & 22.93 & 0.02 & 0.51 & 1.05 & 0.29 & 10.23 & 22.92 & 0.00 & 0.00 & 0.00 & 1.00 & 0.00 \\ 
  In recovery & 0.25 & 0.43 & 0.02 & 0.01 & 0.95 & 0.34 & 0.25 & 0.43 & -0.00 & -0.00 & -0.00 & 1.00 & 0.00 \\ 
  Internal Mental Distress Scale & 8.25 & 8.87 & 0.06 & 0.57 & 3.02 & 0.00 & 8.26 & 8.87 & 0.00 & 0.00 & 0.00 & 1.00 & 0.00 \\ 
  Behavior Complexity Scale & 10.63 & 8.20 & 0.04 & 0.37 & 2.11 & 0.04 & 10.63 & 8.19 & 0.00 & 0.00 & 0.00 & 1.00 & 0.00 \\ 
  Traumatic Stress Scale & 2.14 & 3.42 & 0.05 & 0.17 & 2.32 & 0.02 & 2.13 & 3.42 & 0.00 & 0.00 & 0.00 & 1.00 & 0.00 \\ 
  Emotional Problems Scale & 0.25 & 0.20 & 0.04 & 0.01 & 1.94 & 0.05 & 0.25 & 0.20 & 0.00 & 0.00 & 0.00 & 1.00 & 0.01 \\ 
  Categorization of internal and/or external problems &&&&&&&&&&&&& \\
   \hspace{0.5em} \textit{Neither} & 0.37 & 0.48 & -0.05 & -0.02 & -2.29 & 0.02 & 0.37 & 0.48 & -0.00 & -0.00 & -0.00 & 1.00 & 0.00 \\ 
   \hspace{0.5em} \textit{Both} & 0.33 & 0.47 & 0.06 & 0.03 & 2.97 & 0.00 & 0.33 & 0.47 & 0.00 & 0.00 & 0.00 & 1.00 & 0.00 \\ 
   \hspace{0.5em} \textit{Externalizing Only} & 0.22 & 0.42 & -0.02 & -0.01 & -0.95 & 0.34 & 0.22 & 0.42 & -0.00 & -0.00 & -0.00 & 1.00 & 0.00 \\ 
   \hspace{0.5em} \textit{Internalizing Only} & 0.08 & 0.28 & 0.01 & 0.00 & 0.38 & 0.70 & 0.08 & 0.28 & -0.00 & -0.00 & -0.00 & 1.00 & 0.00 \\ 
  \hline\hline
     \textbf{Effective Sample Size} & \multicolumn{6}{c|}{2229} & \multicolumn{6}{c|}{2063.95} \\
  \hline
\end{tabular}}
    \caption{Summary statistics of covariate information, both without (left section) and with (right section) weighting using the entropy balancing weights. The last column presents the KS-statistics comparing the weighted and unweighted covariate distribution for each covariate.  Each section is organized as follows: mean and standard deviation of the covariate (marginal balance), correlation with the exposure variable (correlation balance), and the regression coefficient and corresponding $t$-statistic and $p$-value obtained by regressing exposure on the covariate indicated (conditional mean balance). The last row of the table represents the effective number of observations without and with weighting.  The table illutrates that the entropy balancing removes the observed correlation among the covariates and the exposure and that they are now conditionally independent given the weights.}
    \label{tab:apptab}
\end{table}

\subsection{Balance results} 
Table \ref{tab:apptab} organizes important features of the data set, such as summaries of the marginal relationships and correlations of covariates with the primary exposure variable (number of treatment sessions), both without and with weighting using the entropy balancing weights (matched on three-moments).  While there are only mild correlations among the covariates and the exposure variable prior to weighting, many of these relationships have regression coefficients that are statistically significant.  The table also demonstrates that the entropy balancing weights provide adequate covariate balance as many of the correlations are small and close to zero and there are no longer relationships present when regressing the exposure variable on each covariate in the weighted data.  Another important result is that the weighted distribution of each covariate is relatively unchanged under this weighting scheme (due to the marginal constraints of the algorithm) as illustrated through the similar weighted and unweighted mean and standard deviations, as well as the KS-statistic comparing the weighted and unweighted distributions.  Finally, there is only a small reduction in effective sample size in performing this analysis, which is not surprising given the mild correlations in the unweighted data.  

\begin{figure}
    \centering
    \includegraphics[width=\textwidth]{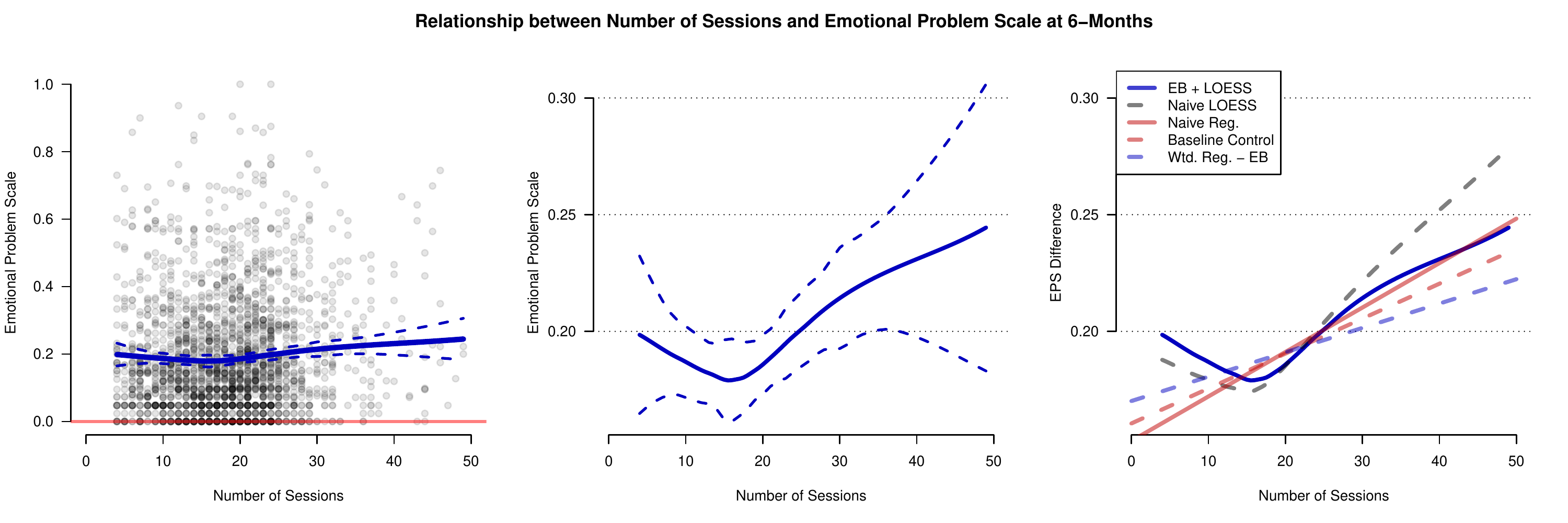}
    \includegraphics[width=\textwidth]{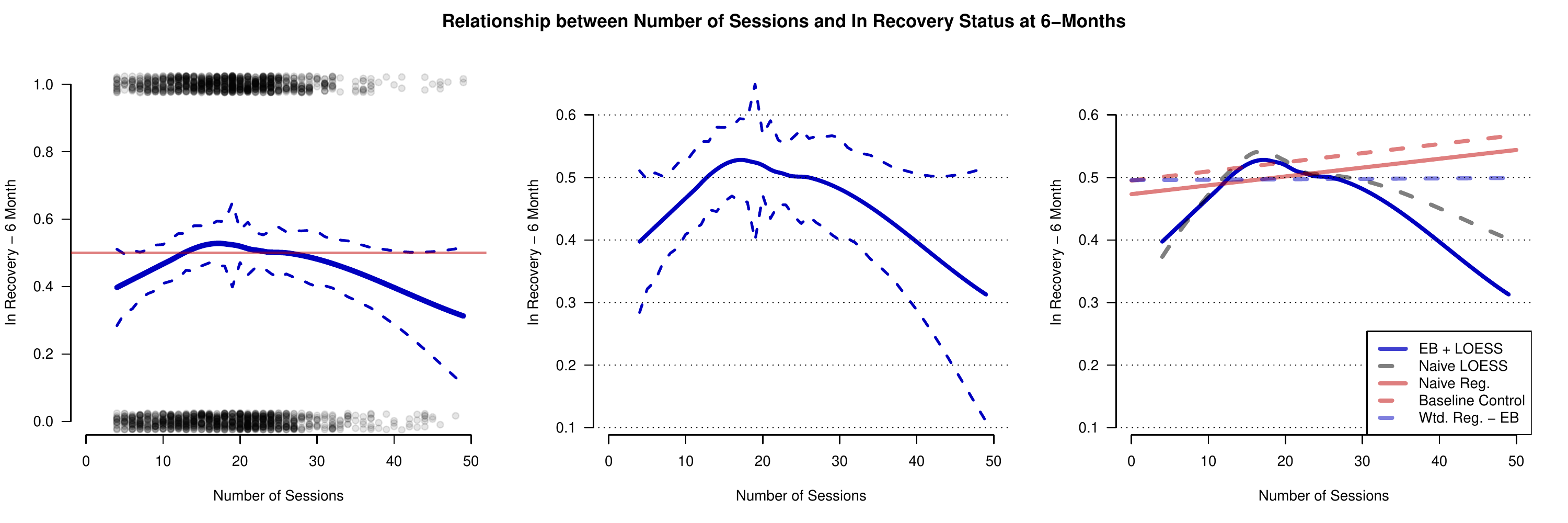}
    \caption{Estimated relationships between the number of treatment sessions and the emotional problem scale (EPS) at the 6-month follow-up (top) and ``in recovery'' status at the 6-month follow-up (bottom).  The right panel of each row compares estimates for the proposed method (EB + LOESS) to 1) a naïve (unweighted) local linear regression, 2) a naïve (unweighted) regression modeled such that exposure is the only covariate, i.e.,  \texttt{Y $\sim$ A}, 3) a simple linear regression controlling for the baseline value of the outcome measure, i.e., \texttt{Y $\sim$ A + X}, and 4) a weighted linear regression only controlling for the exposure variable, i.e., \texttt{Y $\sim$ A}.}
    \label{fig:appfig2}
\end{figure}

\subsection{Outcome analysis:  Dose-Response Curves}
Figure \ref{fig:appfig2} presents the results from performing weighted local linear regressions using the entropy balancing weights.  The left panels show the estimated curve and the bootstrapped 95\% confidence interval overlain onto the original data and the center panels focus in on the relationships.  The right panels provide a closer look at the curves and compares against other methods of estimating the outcome variables.  Specifically, we demonstrate the results using a naïve (unweighted) local linear regression, a naïve (unweighted) regression, i.e., $Y \sim A$, a regression controlling for baseline variables of interest (EPS or ``in recovery'' status at baseline), and a weighted regression using the entropy balancing weights and a simple model, i.e., $Y\sim A$. Comparing across these results, we see that our method provides results that are somewhat distinct from the other methods of estimation and that the results for the in recovery variable and the EPS are similar (but trending in opposite directions). Combined with the simulation results of the last section, this suggests that the method is able to more accurately estimate the true marginal relationship and that the use of nonparametric estimation frees the analyst from having to determine the correct functional relationship.  

These results also provide a few interesting insights into the relationship between A-CRA and our outcomes.  The first result is that there is a positive approximately quadratic relationship between the exposure variable and EPS at 6-months; those with the lowest and highest number of treatment sessions appear to have ``worse'' outcomes on EPS. Similarly, there is a negative approximately quadratic relationship between the number of treatment sessions and in recovery status; those exposed to fewer sessions do similar to those with many sessions, and both perform worse on this key substance use outcome than those toward the center of the exposure distribution. Taken causally, these results would be confusing because it appears that the number of treatment sessions causes worse outcomes on the EPS and substance use outcomes, when a monotonic relationship may be expected if the number of sessions is playing a role in these outcomes.  As was discussed earlier, the number of sessions likely violates some of the key assumptions for a causal interpretation and a more reasonable interpretation is that there is an interaction between the clinician and the client throughout the A-CRA exposure that explains these conflicting results.  That is, it could be that individuals receiving fewer sessions are not truly engaging in treatment and this does not provide enough ``dose'' in the form of number of treatment sessions to the client. Subsequently, they perform worse on the substance use outcome of being in recovery.  Similarly,  those that have a larger number of sessions are more difficult to treat, and therefore we see worse outcomes at these levels.  Compared with those two groups, those in the middle of the exposure distribution are more ``typical'' cases and we see these individuals having the best outcomes. This essentially suggests a form of unobserved confounding and highlights the need for well-defined treatments and critical thinking about an ideal experimental set-up that one would want to perform to address the scientific question.  The lack of a rigorous causal interpretation does not diminish that these are interesting results that provide useful insights to clinicians, such as the fact that more information should be recorded and considered at intake on the individuals and throughout their substance use program to better understand and act upon the underlying differences among patients at different exposure levels. 

\section{Discussion} \label{sec:discussion}

This manuscript outlined a nonparametric estimation procedure using local linear regression that incorporates entropy balancing weights for performing inference in observational settings.  The method was demonstrated through a comprehensive simulation study and through an applied example that assessed the relationship between the number of treatment sessions received during substance use treatment on substance use and emotional well-being outcomes among adolescent substance users.  The simulation study demonstrated that the methodology can provide comparable (or in some cases superior) performance to a competing method (CBPS) and that both out perform naïve analyses using linear models.  Additionally, the number of moments that are balanced in the analysis is critical.  When only the first moment was balanced, the methods did similar to naïve analyses; including higher moments in the balancing equations further reduces the bias in the analysis for both entropy balancing and the CBPS.  The application demonstrated that the method can be useful in health care settings and provide insight to clinicians, while freeing them from modeling decisions in the estimation phase (though it does not free them from critically thinking about assumptions for causal inference and underlying data generation).  

Causality is often the underlying intention (albeit in some cases a subconscious intention) when collecting data in an attempt to organize phenomenon in the world around us and requires a fine control on many conditions in the world.  Establishing ``true'' causality is often unattainable though (see the \textit{Fundamental Problem of Causal Inference}) and we must make assumptions that enable us to find comparisons of outcomes in the world under counterfactual (i.e., alternative) exposures.  The method developed here is no different and requires many strong assumptions that can be implausible and in some cases untestable.  The method supposes that a suitable estimand in an observational study is the average dose-response effect within a population, an analog from clinical trials, that provides insight on the broader behavior of an outcome as it relates to the exposure.  Researchers should be cautioned though that the positivity assumption is necessary to defend this estimand and it requires that for each value $\bs{X}=\bs{x}$ there is a positive probability of receiving \textbf{any} exposure level in the range of exposures of interest (something guaranteed in a completely randomized experiment) in order to enable estimation. However, in many observational settings, this can be an unrealistic assumption.  Any form of moderate preferable selection of exposures that can be ascribed to covariates, both observed and unobserved, makes satisfying this assumption less plausible.  When there are strong relationships in the data, then estimation of the \textit{conditional or local average exposure effect} should be considered by thinking through the suitability of average effects in some neighborhood of $\bs{X}=\bs{x}$.  One diagnostic for assessing the positivity assumption in observational settings is assessing the effective sample size.  When this measure tends toward small numbers, there is not enough sample to meet the constraints within the balancing algorithm and implies that the population dose-response curve estimand should not be addressed with the data.  

Even when a positivity assumption appears plausible, there are additional hurdles to making causal claims that a researcher must consider such as temporal ordering, the structure of the relationships between covariates and the exposure, unconfoundedness, and model selection.  Our methodology attempts to relieve the researcher of one of these (i.e., model selection) by using local linear models combined with entropy balancing, but the researcher must defend and document assertions that the remaining assumptions hold.  In particular, in mental health and health care settings, both the structure of the relationships among variables and the unconfoundedness assumption must be made with care.  In our example, it would appear that temporal ordering and positivity are plausible, but there appears to be some unobserved variables in the relationship between the number of sessions and the outcomes due to the longitudinal nature of the decision process in ``assigning'' treatment.  These deficiencies do not imply that the application of these methods do not inform science, but they should come with the appropriate caveats that this work mainly informs hypothesis generation and recommendations for future work.  A related topic within the field of causal inference are sensitivity analyses that attempt to isolate how sensitive a result is to assumptions made within the design pipeline.  Future work should address and develop sensitivity analyses in observational studies with continuous exposures in nonparametric modeling settings.  

The methods developed here provide a unique flexibility for estimation in observational settings that incorporates weights to remove biases.  While the assumptions required to make a causal claim are numerous, the combination of nonparametric estimation and entropy balancing weights can clearly remove many biases in observational studies and should become a common tool for applied researchers.  

\section*{Acknowledgements}

The authors would like to thank many researchers that helped us to think critically about these ideas.  Specifically, Lane Burgette, Megan Schuler, and Joseph Pane for their advice and support in finishing this draft.  Additionally, we would like to thank our collaborative partners who have provided insights during our understanding of the applied example, specifically, Mark Godley, Lynsay Ayer, and Rod Funk.   

\funding{Research reported in this manuscript was supported by the National Institute on Drug Abuse of the National Institutes of Health under award number R01DA045049. The content is solely the responsibility of the authors and does not necessarily represent the official views of the National Institutes of Health.}

\data{The development of this manuscript was supported by the Center for Substance Abuse Treatment (CSAT), Substance Abuse and Mental Health Services Administration (SAMHA) [\#270-2003-00006, \#270-2003-00006, and \#270-2007-00004C] using data provided by the following grantees: TI-15413, TI-15415, TI-15421, TI-15433, TI-15438, TI-15446, TI-15447, TI-15458, TI-15461, TI-15466, TI-15467, TI-15469, TI-15475, TI-15478, TI-15479, TI-15481, TI-15483 TI-15485, TI-15486, TI-15489, TI-15511, TI-15514, TI-15524, TI-15527, TI-15545, TI-15562, TI-15577, TI-15584, TI-15586, TI-15670, TI-15671, TI-15672, TI-15674, TI-15677, TI-15678, TI-15682, TI-15686, TI-17589, TI-17604, TI-17605, TI-17638, TI-17646, TI-17648, TI-17673, TI-17702, TI-17719, TI-17728, TI-17742, TI-17744, TI-17751, TI-17755, TI-17761, TI-17763, TI-17765, TI-17769, TI-17775, TI-17779, TI-17786, TI-17788, TI-17812, TI-17817, TI-17821, TI-17825, TI-17830, TI-17831, TI-17847, TI-17864, TI-20759, TI-20781, TI-20798, TI-20806, TI-20827, TI-20828, TI-20847, TI-20852, TI-20865, TI-20870, TI-20910, TI-20946, TI-23174, TI-23186, TI-23188, TI- 23195, TI-23196, TI-23197, TI-23200, TI-23202, TI-23204, TI-23206, TI-23224, TI-23244, TI-23247, TI-23265, TI-23270, TI-23276, TI-23278, TI-23279, TI-23296, TI-23298, TI-23304, TI-23310, TI-23311, TI-23312, TI-23316, TI-23322, TI-23323, TI-23325, TI-23336, TI-23345, TI-23346, TI-23348. The authors thank these agencies, grantees, and their participants for agreeing to share their data to support this secondary analysis. The opinions about these data are those of the authors and do not reflect official positions of the government or individual agencies. Please direct correspondence to Brian Vegetabile, bvegetab@rand.org.}

\bibliographystyle{agsm}

\bibliography{ebcontinuous.bib}

\appendix

\section{Binary Exposure Entropy Balancing Algorithm - Hainmueller (2012)} \label{app:binary}

Consider an outcome variable $Y$, a binary exposure variable $A$, a multivariate vector of covariates $X$, and a distribution represented by $p(x)$ over which inference is required with specified moments $\mu_j^p = E_{X}[X_j^p] \equiv \int_{\mathcal{X}} x_j^p p(x)\partial x$.  The goal in the binary setting is estimating weights $w_i, w_i \ge 0, \sum_i w_i = 1$ such that the estimator in Equation (\ref{eqn:wtdest}), i.e., differences in potential outcomes, has consistent estimation for $\tau = E_{X}[Y(1)-Y(0)]$.  
\begin{equation}
    \hat \tau = \sum_{i=1}^N w_i I(A_i = 1) Y_i - \sum_{i=1}^N w_i I(A_i = 0) Y_i \label{eqn:wtdest}
\end{equation}

Entropy balancing selects the weights such that the weighted empirical moments within each group $A=a$ match those from the population of interest, i.e., 
\begin{align*}
    \mu_j^p \approx \sum_{i=1}^N w_i I(A_i = a) X_{ij}^p
\end{align*}

To do this it minimizes the KL-divergence between the distribution of the weights that satisfies the desired distributional characteristics (represented by constraints on the weighted moments) and a ``base'' distribution of weights $q$ (typically defined as $q_i \equiv N^{-1}$ representing the empirical weights) within each group:
\begin{center}
\begin{tabular}{lr}
    $\displaystyle\min_{w} \sum_{i : A_i = a} w_i \log \frac{w_i}{q_i}$ & \hspace{1em} such that  \\
    $\displaystyle\sum_{i: A_i = a} w_i X_{ij}^p = \mu_{j}^p$ & \hspace{1em} for all $j,p$ \\ 
    $\displaystyle \sum_{i : A_i = a} w_i = 1, w_i \ge 0$
\end{tabular}
\end{center}

\cite{hainmueller2012entropy} chose to optimize this object using the method of Lagrange multipliers,

\begin{align*}
    \min L_p = \sum_{i : A_i = a} w_i \log \frac{w_i}{q_i} + \sum_{j,p} \lambda_{j.p} (w_i X_{ij}^p - \mu_{j}^p) + (\lambda_0 - 1) \left(\sum_{i : A_i = a} w_i - 1\right)
\end{align*}

The objective above provides a closed form solution for the weights that can dramatically reduce the dimensionality of the optimization problem and facilitate finding an efficient solution using off-the-shelf optimizers.  See \cite{hainmueller2012entropy} for a full treatment of these methods in the case of binary treatments for estimating the average effect on the exposed.

\section{Supplemental Simulation - No Effect of Exposure}

In this supplemental material, we reconsider the simulation of Section \ref{sec:simulation}, but modify the outcome variable so that there is no direct effect of the exposure variable on the outcome variable $Y$.  To accomplish this, in this simulation the data generation procedure is augmented such that
\begin{align*}
    Y &= X_1 + X_1^2 + X_2 + X_2^2 + X_1 X_2 + X_3 + \epsilon, \hspace{1em} \\
    \epsilon &\sim \mathcal{N}(0,1)
\end{align*}
and all other data-generating parameters remain the same.  This implies that the marginal effect is $E[Y(a)] = 3.55$ for all $a \in \mathcal{A}$ and thus $A$ has no population-level effects.  Within the simulation we again model using the local regression method, but here compare against a global linear model where the assumption is that $Y(a) \approx \beta_0 + \beta_1 A$ to assess how the weights perform without requiring a selection of the correct form of the outcome model.  

\begin{figure}[H]
    \centering
    \includegraphics[width=\textwidth]{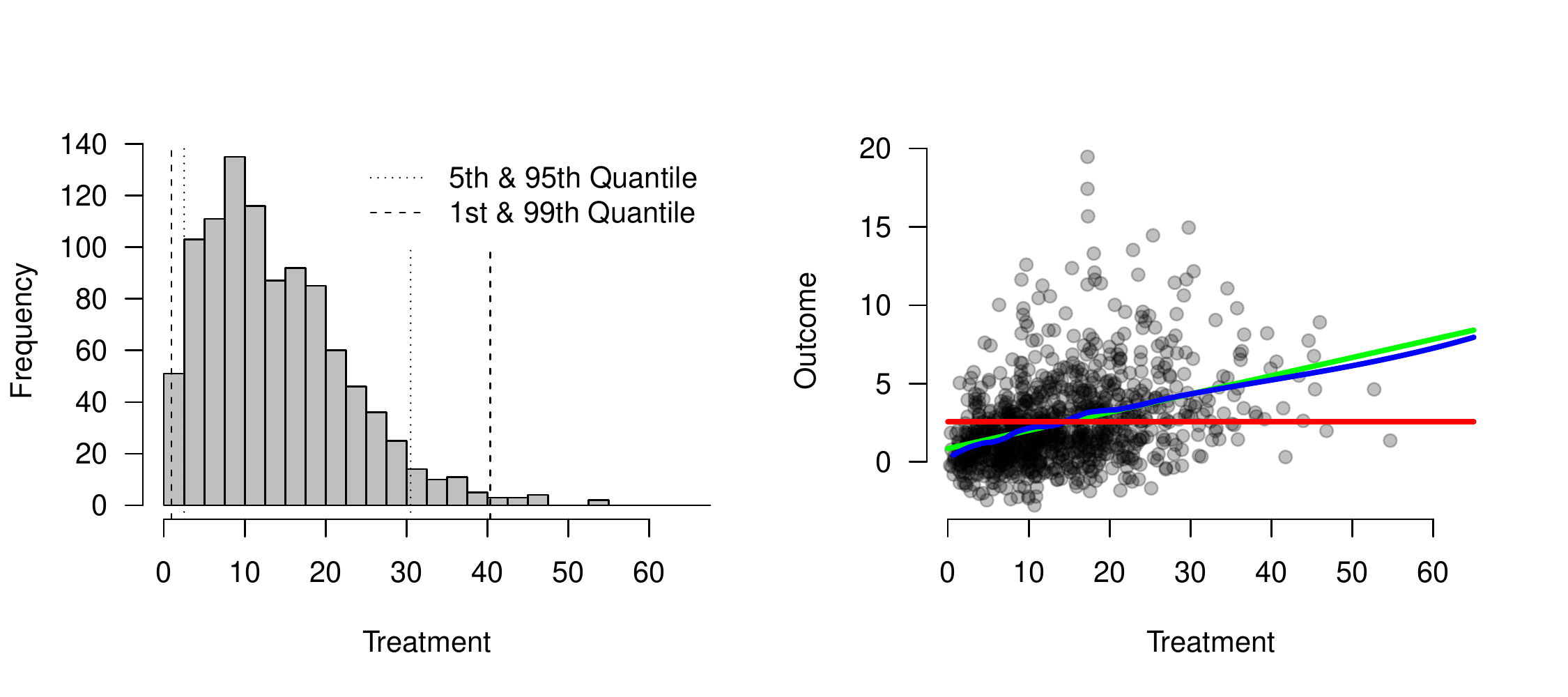}
    \caption{Visualization of the distribution of treatment variable (left) and relationship between treatment and outcome (right).  In the left figure, the high-density region of $A$ is highlighted.  In the right figure, the true marginal relationship is shown in {\color{red} \textbf{red}}, a simple unweighted linear smoother is shown in {\color{blue} \textbf{blue}}, and an unweighted linear estimation is provided in {\color{green} \textbf{green}}.}
    \label{fig:fig1noeffect}
\end{figure}

A visualization of the induced relationship between exposure and outcome is shown in Figure \ref{fig:fig1noeffect}. 

\subsection{Results}

\subsubsection{Covariate Balance}

The covariate balance results are unchanged as the relationships among $A$ and $X$ in the data-generation are the same and the seed used for the random number generator is the same.

\subsubsection{Performance in Estimating Dose-Response Curves}

The performance here is similar to the results of Section \ref{sec:simulation}.  Both entropy balancing and the CBPS perform well when a high enough number of moments is balanced and all other methods perform less than satisfactory.  The bootstrap confidence intervals again remain conservative, but in this case it is observed that at certain points of the exposure distribution there is under-coverage.  Future work should further investigate confidence intervals for these methods.  

\begin{table}[H]
    \centering
    \begin{tabular}{l|c|c|c|c|c}
    && \multicolumn{2}{c|}{Average Bias} & \multicolumn{2}{c}{Mean Square Error}\\\hline
    Weighting Method & Avg. ESS & LOESS & Reg. & LOESS & Reg. \\\hline 
  Unweighted & 1000.000 & 1.129 & 1.196 & 6.031 & 6.347 \\ \hdashline
  Entropy Balancing (1) & 737.228 & 0.812 & 0.907 & 3.241 & 3.142 \\ 
  Entropy Balancing (2) & 463.829 & -0.102 & 0.093 & 0.359 & 0.047 \\ 
  Entropy Balancing (3) & 406.130 & -0.047 & 0.130 & 0.381 & 0.059 \\ 
  Entropy Balancing (4) & 385.312 & -0.093 & 0.123 & 0.446 & 0.057 \\ \hdashline
  Linear Model & 137.388 & 0.635 & 0.627 & 4.172 & 3.558 \\ \hdashline
  CBPS - Nonparametric: (1) & 441.807 & 0.953 & 1.017 & 3.998 & 3.557 \\ 
  CBPS - Nonparametric: (2) & 243.853 & -0.056 & 0.049 & 0.628 & 0.059 \\ 
  CBPS - Nonparametric: (3) & 219.252 & -0.006 & 0.039 & 0.756 & 0.064 \\ 
  CBPS - Nonparametric: (4) & 225.603 & -0.027 & 0.044 & 0.764 & 0.074 \\ 
    \end{tabular}
    \caption{Results in estimating the population dose-response curve when there is no relationship between exposure and outcome.}
    \label{tab:my_label}
\end{table}

\begin{figure}[ht!]
    \centering
    \includegraphics[width=\textwidth]{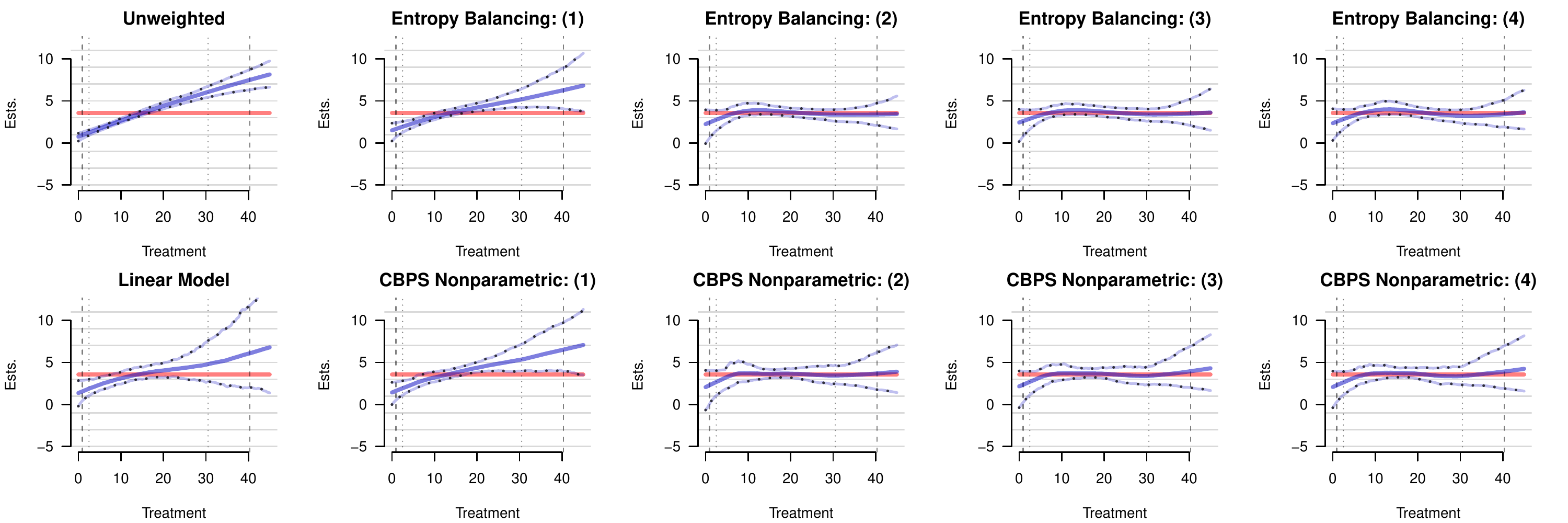}
    \caption{Performance estimating the dose-response curve across repeated simulated samples: Local Linear Regression}
    \label{fig:response1noeffect}
\end{figure}

\begin{figure}[ht!]
    \centering
    \includegraphics[width=\textwidth]{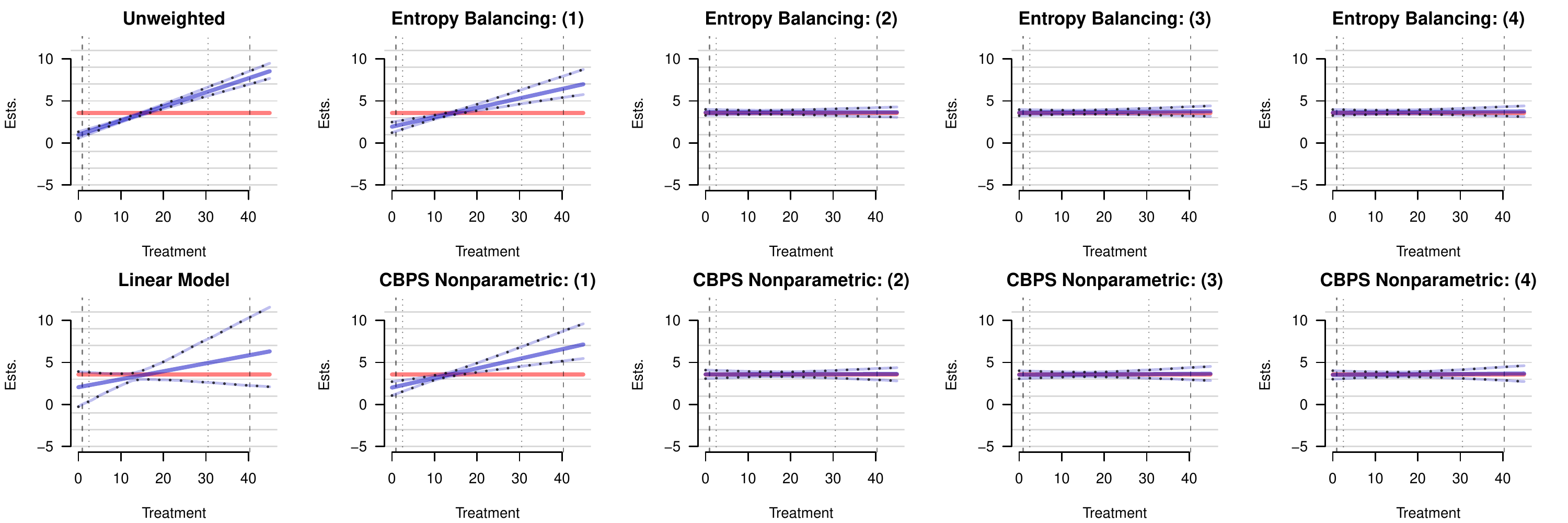}
    \caption{Performance estimating the dose-response curve across repeated simulated samples: Linear Regression - outcome is modeled using linear relationship with exposure}
    \label{fig:response2noeffect}
\end{figure}

\begin{figure}
    \centering
    \includegraphics[width=\linewidth]{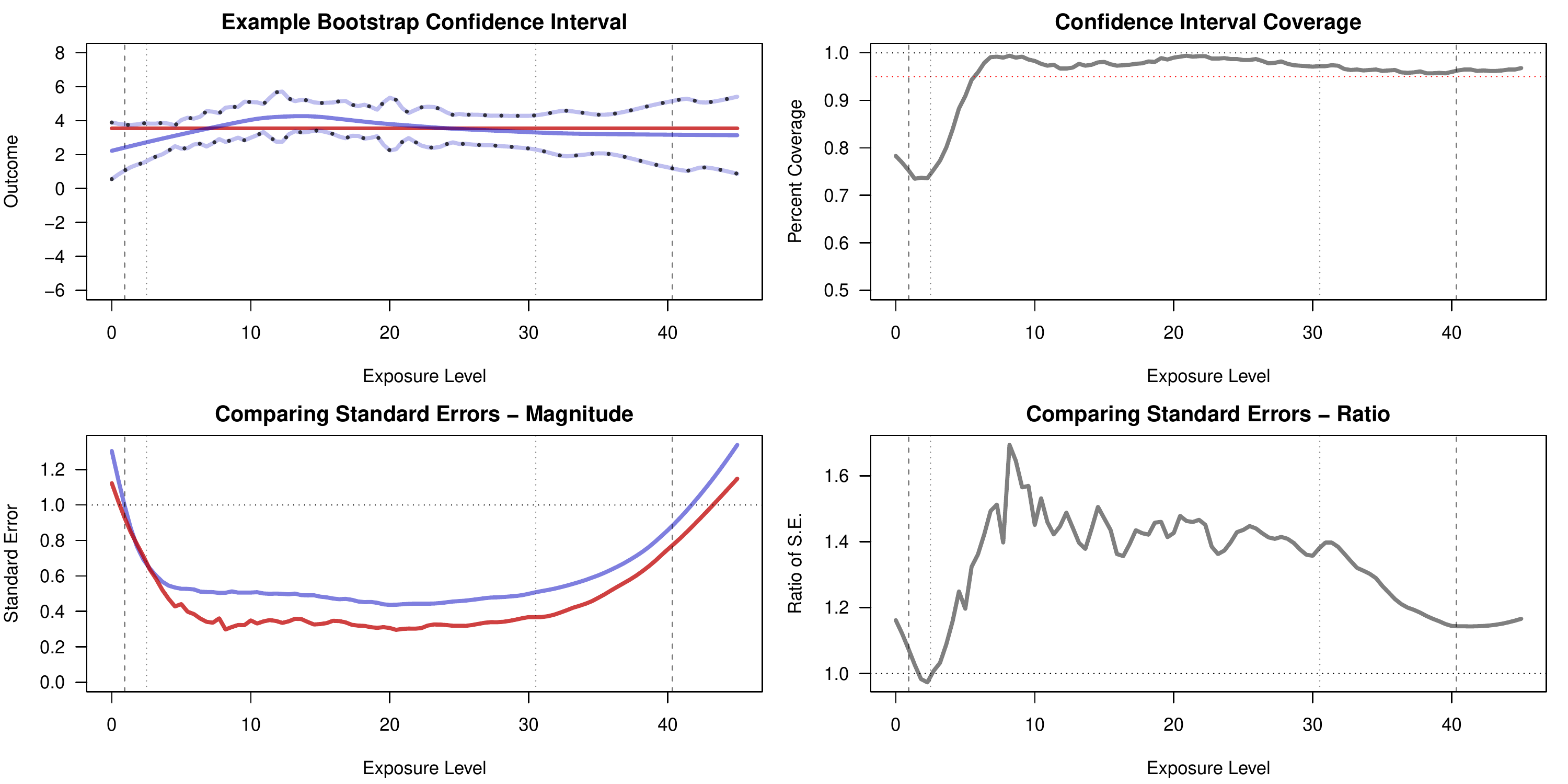}
    \caption{Simulation results for bootstrap confidence intervals. Upper left panel contains a single visualization of a bootstrapped confidence interval for the curve. Upper right panel contains the point-wise coverage of the 95\% bootstrapped confidence intervals. Lower panels contain the magnitude (left) and ratio (right) of the average point-wise bootstrap standard error across the bootstrap simulations as compared to the standard error of the estimated curves obtained in this simulation.  In all figures the vertical lines represent the $1^{st}, 5^{th}, 95^{th}, 99^{th}$ quantiles of the distribution of $A$ in the high-density region.}
    \label{fig:bootstrap}
\end{figure}

\end{document}